\documentclass[a4paper,11pt]{article}
\pdfoutput=1 

\usepackage{jcappub} 

\usepackage{natbib}
\usepackage{aas_macros}
\usepackage[T1]{fontenc} 
\usepackage{graphicx}	
\usepackage{amsmath}	
\usepackage{amsfonts}   

\usepackage{amssymb}	
\usepackage{mathrsfs}
\usepackage{changepage}
\usepackage{import}
\usepackage{wrapfig}

\usepackage{footnote}
\usepackage[bottom]{footmisc}
\usepackage{subcaption}
\usepackage{booktabs}
\usepackage{graphics}
\usepackage{adjustbox}
\usepackage{blindtext}
\usepackage{tabularx}
\usepackage{url}
\usepackage{breakurl}
\usepackage{fontawesome}

\DeclareUnicodeCharacter{2009}{\,}
\DeclareUnicodeCharacter{223C}{\,}
\usepackage{xcolor, soul}
\definecolor{deep21green}{HTML}{c6e2c7}
\sethlcolor{deep21green}

\def\deeep{\texttt{deep21}}

\def\nhat{\hat{\textbf{n}}}

\def\kpara{{k_{\parallel}}}

\def\cosmo{\text{cosmo}}
\def\fgd{\text{fg}}

\usepackage{tikz,xcolor,hyperref}

\definecolor{lime}{HTML}{A6CE39}
\DeclareRobustCommand{\orcidicon}{%
	\begin{tikzpicture}
	\draw[lime, fill=lime] (0,0) 
	circle [radius=0.16] 
	node[white] {{\fontfamily{qag}\selectfont \tiny ID}};
	\draw[white, fill=white] (-0.0625,0.095) 
	circle [radius=0.007];
	\end{tikzpicture}
	\hspace{-2mm}
}

\foreach \x in {A, ..., Z}{%
	\expandafter\xdef\csname orcid\x\endcsname{\noexpand\href{https://orcid.org/\csname orcidauthor\x\endcsname}{\noexpand\orcidicon}}
}

\graphicspath{{./}{figures/}}

\title{\texttt{deep21}: a Deep Learning Method for 21cm Foreground Removal}

\author[a,b,c,1]{T. Lucas Makinen,\orcidA{}\note{Corresponding author.}}
\author[a]{Lachlan Lancaster,\orcidB{}}
\author[a]{Francisco Villaescusa-Navarro,\orcidC{}}
\author[a,d]{Peter Melchior,\orcidD{}}
\author[e]{Shirley Ho,}
\author[e,f,g]{Laurence Perreault-Levasseur,\orcidE{}}
\author[e,a]{and David N. Spergel}

\affiliation[a]{Department of Astrophysical Sciences, Princeton University, \\Peyton Hall, Princeton, NJ, 08544, USA}
\affiliation[b]{Institut d'Astrophysique de Paris, Sorbonne Universit\'e, \\98 bis Boulevard Arago, 75014 Paris, France}
\affiliation[c]{Center for Statistics and Machine Learning, Princeton University, \\Princeton, NJ 08544, USA}
\affiliation[d]{Center for Computational Astrophysics, Flatiron Institute, \\162 5th Avenue, New York, NY, 10010, USA}
\affiliation[e]{Department of Physics, Univesit\'e de Montr\'eal,  \\CP 6128 Succ. Centre-ville, Montr\'eal, H3C 3J7, Canada}
\affiliation[f]{Mila - Quebec Artificial Intelligence Institute, \\Montr\'eal, Canada}

\emailAdd{timothy.makinen@cfa.harvard.edu}
\emailAdd{lachlanl@princeton.edu}
\emailAdd{fvillaescusa-visitor@flatironinstitute.org}
\emailAdd{melchior@astro.princeton.edu}
\emailAdd{shirleyho@flatironinstitute.org}
\emailAdd{dspergel@flatironinstitute.org}

\abstract{
We seek to remove foreground contaminants from 21cm intensity mapping observations. We demonstrate that a deep convolutional neural network (CNN) with a UNet architecture and three-dimensional convolutions, trained on simulated observations, can effectively separate frequency and spatial patterns of the cosmic neutral hydrogen (HI) signal from foregrounds in the presence of noise. Cleaned maps recover cosmological clustering amplitude and phase within 20\% at all relevant angular scales and frequencies. This amounts to a reduction in prediction variance of over an order of magnitude across angular scales, and improved accuracy for intermediate radial scales ($0.025 < \kpara{} < 0.075\ \rm h\ Mpc^{-1})$ compared to standard Principal Component Analysis (PCA) methods. We estimate epistemic confidence intervals for the network's prediction by training an ensemble of UNets. Our approach demonstrates the feasibility of analyzing 21cm intensity maps, as opposed to derived summary statistics, for upcoming radio experiments, as long as the simulated foreground model is sufficiently realistic.
We provide the code used for this analysis on \href{https://github.com/tlmakinen/deep21}{GitHub \faGithub}, as well as a browser-based tutorial for the experiment and UNet model via the accompanying \href{http://bit.ly/deep21-colab}{Colab notebook \faGoogle}. 
}

\keywords{cosmology: radio -- reionisation, large-scale structure -- foregrounds -- deep learning, signal processing }

\begin{document}
\maketitle
\flushbottom

\section{Introduction} \label{sec:intro}

Observations of cosmic neutral hydrogen emission hold the promise of
making precision measurements of the universe's evolution at intermediate 
to late redshifts ($0.5 > z > 30$) \citep{Madau97,Tozzi00,LiuShaw20,PUMA}, providing an observable to trace both the growth of massive structure from the time of the Cosmic Microwave Background (CMB), as well as constrain the physics of Reionization ($10 > z > 6$) \citep{Furlanetto06, PritchardLoeb12, MoralesWyithe10}. 

Measurement of the 21cm line relies on intensity mapping, in which large fractions of the sky are observed to capture wide-field statistics, instead of resolving individual sources \citep[for an overview see e.g.][]{MoralesWyithe10, LiuShaw20}. This technique has already been used to place constraints on the formation of the first stars and galaxies at $z \sim 9$ \citep{edges2017}, but the precise nature of the Epoch of Reionization (EoR) remains unknown \citep{Furlanetto06}. 21cm intensity maps also promise to be a tracer of three-dimensional, large-scale structure growth at later redshifts $z \lesssim 4$, linking late-stage structure to underlying gravitational theory and the primordial density \citep{hall-2013, camera-2013}. Upcoming experiments, such as the Square Kilometer Array (SKA) promise to trace EoR physics to large-scale structure formation using this single observable \citep{ska2020}.

The greatest challenge for these measurements is mitigating systematics and 
removing enormous foreground contamination from galactic radio sources such as synchrotron and free-free emission, as well as 
extragalactic features like point sources \citep{MoralesWyithe10, PritchardLoeb12}.  These contaminants tend to be three to 
four orders of magnitude brighter than the interesting cosmological signal 
\citep{haslam1982, MoralesWyithe10}. Furthermore, foregrounds lack detailed analytic descriptions, making 21cm likelihoods hard to specify \citep{Alonso_sim_2014}.

The foreground signals have different statistical properties than the 
cosmological signal, a phenomenon thoroughly covered in the literature \citep{Di_Matteo_2002, Peng_Oh_2003, santos-fg, wang_poly_2006, Morales_2006, Jeli__2008, Gleser_2008, Bernardi_2009, Bernardi_2010, Moore_2013}, with several proposed methods for signal separation \citep[][]{Liu_2009, Wolz_2014, Liu_2011, Masui_2013, Shaw_2014, Shaw_2015}.
Most foreground contaminants are forecast to be spectrally smooth in frequency, motivating the application of blind signal separation 
techniques, such as Principal Component Analysis (PCA)
\citep[e.g.][]{pca_oliveira2008,Alonso_pca_2014}, which require no prior 
knowledge of the expected signals. However, blind separation techniques 
are not linked to physical processes and therefore make no use of our 
physical understanding of these foregrounds. This means that there exists 
information in the observed signal that is not fully exploited for 
separation. Blind subtraction for single-dish experiments irretrievably removes the mean of the HI intensity spectrum and can also distort the signal anisotropy \citep{Cunnington_2019, 2015ApJ_signal_loss_singledish, 2018ApJ_signalloss_inter}, placing the focus of analyses on interpreting compressed,
sometimes biased \citep[see e.g.][]{Wolz_2014, Spinelli_2019, Alonso_pca_2014}, summary
statistics derived from these maps, such as power spectra. With clean 21cm intensity
maps, more fundamental large-scale structure analyses and parameter extraction would be
possible using the maps themselves \citep[][for a review]{deeplearn-eor-Gillet_2019, Mangena_2020, Weltman_2020}.

We address these problems by constructing a convolutional neural network 
to recover cosmological 21cm maps from PCA-reduced inputs. Deep learning
architectures are ideally suited to similar high-dimensional problems such 
as image segmentation, classification, and computer vision tasks (see e.g. 
\cite{GoodBengCour16} for a review, and e.g. \cite{shirleyd3m}, 
\cite{prob-unet}, \cite{Jay_2020} for applications), in which patterns and higher-order 
correlations must be captured over a large set of input data.  To 
incorporate an estimate of uncertainties of the separated maps we train an 
ensemble of networks on a suite of simulated radio skies. We then test 
these architectures on simulations with altered foreground parameters to 
assess how well the approach generalizes beyond the fiducial choice of model.

{Recent studies have incorporated deep learning techniques to analyze EoR 
cosmology \citep{deeplearn-eor-Gillet_2019, deeplearn-21-Mangena_2020,
Kwon-deeplearn-21_2020, Pablo_2020, Hassan2020, Chardin_2019, List_2020}, but have largely focused on retrieving compressed cosmological statistics or higher-order correlations from clean 
intensity maps.
Some novel methods for astrophysical foreground removal include \citet{bayesian-semi-blind}, who present a Bayesian method for power spectrum recovery that effectively limits blind subtraction bias by exploiting the isotropy and homogeneity of the 21cm signal. Deep learning foreground removal techniques have also been investigated, e.g. for CMB maps \citep{yao_2018}, interferometric 21cm foregrounds \citep{Li_2019}, and far-field radio astrophysics \citep{Pablo_2020} .}

However, our study is (to our knowledge) the first to leverage a fully 3D UNet architecture to separate clean 21cm maps from radio foregrounds directly from simulated single-dish observations. These 
clean maps can then be leveraged in existing 21cm analyses.

The layout of this paper is as follows: in Section \ref{sec:formalism}, we 
present the physical formalism behind HI intensity mapping and astrophysical 
foregrounds. In Section \ref{sec:methods}, we present the foreground 
subtraction techniques we employ to train our network. We detail the 
architecture choice and the UNet training procedure in Section \ref{sec:deep21}. 
Results for both blind subtraction and our network are presented in Section 
\ref{sec:results}, followed by tests on foregrounds with altered simulation 
parameters. We discuss the successes of our network, as well as failure modes 
in Section \ref{sec:conclusions}. The cosmology we assume in our study is the 
standard flat $\Lambda$CDM in agreement with the results from the \citet{planck2016}, 
with fiducial parameters $\{ \Omega_m, \Omega_b, h, n_s, \sigma_8 \}=$ $\{ 0.315, 
0.049,$ $0.67, 0.96, 0.83 \}$.

\section{Methods and Formalism}\label{sec:formalism} 

In this section we describe the theoretical formalism behind the three main 
components of the observed HI 21cm sky: the cosmological signal itself, the 
various galactic and extra-galactic foregrounds, and observational noise. 
For a deeper discussion of various aspects of these components, we refer 
the reader to one of the several review papers on the subject such as 
\cite{Furlanetto06,MoralesWyithe10,PritchardLoeb12} and \cite{LiuShaw20}. 
We then describe how these various components are created in the simulated 
skies that we use to train our machine learning framework.

\subsection{Cosmological HI Signal}
The component of the 21cm sky that we care about most is the redshifted 
HI signal itself. This signal is often described in terms of a brightness 
temperature, $T_b$, which relates the observed intensity of the signal at 
a given sky position and frequency to a temperature \citep{field_58, field_59}. In 
the Rayleigh-Jeans limit ($\hbar \nu_{21} \ll k_B T_b$), this brightness 
temperature can be related to the underlying cosmology at a given line of 
sight $\nhat$, and frequency, $\nu$ as \citep{ Madau97}
\begin{equation}\label{eq:temp-from-density}
    T_b(\nhat, \nu) =  \frac{3 \hbar c^3 A_{21}}{16 k_B \nu^2_{21}}    \frac{(1+z)^2}{H(z)}  {n}_{\rm HI}(z, \nhat)  \, ,
\end{equation}
where $n_{\rm HI} \propto (1 + \delta_{\rm HI})$ is the comoving number 
density of HI, $H(z)$ is the Hubble parameter as a function of redshift, 
$z$, $k_B$ is Boltzmann's constant, $\nu_{21}$ is 
the frequency associated with the HI fine-structure line, $\hbar$ is the 
reduced Planck's constant, and $A_{21} = 2.876\times10^{-15}\ \rm Hz$ is 
the 21cm line Einstein emission coefficient. Using the standard values 
for the various constants along with a standard flat $\Lambda$CDM cosmology for the evolution 
of $H(z)$, this expression can be written in terms of the HI overdensity, 
$\delta_{\rm HI}=\rho_{\rm HI}/\bar{\rho}_{\rm HI}-1$, redshift, and 
cosmological parameters \citep{Madau97, Furlanetto06} as
\begin{equation}\label{eq:temp-hi-density}
\begin{split}
    T_{b}&(\nhat, z) ={} 0.19055\ \times  \\
    &\frac{\Omega_b h (1 + z)^2 x_{\rm HI}(z)}{\sqrt{\Omega_m(1+z)^3 + \Omega_{\Lambda}}} (1 + \delta_{\rm HI}\left(\nhat, z \right))~\text{mK}
\end{split}
\end{equation}
where $h\equiv H_0/(100 \, {\rm km/s/Mpc})$ is the dimensionless Hubble 
constant, $x_{\rm HI}$ is the fraction of baryonic mass comprised of HI, 
and $\Omega_b$ and $\Omega_m$ are the baryon and total matter fractions, 
respectively. The observed 21cm signal can thus be related to the underlying 
cosmological model and relevant parameters for inference studies
\citep{Abdalla_2005, scott-dark-ages, bull-late-stage-cosmo-21, paco-2016, Paco_2018}. The 21cm signal 
can thus be used as a tracer of the large-scale structure of the Universe.

The \texttt{CRIME} simulation code, described in \cite{Alonso_sim_2014}, 
generates a dark matter field and then utilizes a log-normal model to 
generate the cosmological HI intensity maps. Concisely, Gaussian density 
and velocity perturbations are generated on a Cartesian grid with no 
redshift effects. The ``observer'' is placed at the center of the grid, 
and the signal is projected onto the observer's light cone. The Gaussian 
density field then undergoes localized log-normal transformations to 
generate the non-uniform HI density field. The results are projected onto 
spherical sky maps at different frequencies corresponding to the redshift of the structure from the observer, using the \texttt{HEALPix} 
pixelization scheme \citep{healpix}. The brightness temperature $T_b$ is 
related to the underlying HI number density $n_{\rm HI}$, as shown in 
Equation \ref{eq:temp-from-density}. The simulations are generated in a box 
of size 8850 $h^{-1}$ Mpc per side with 3072$^3$ box cells. The interpolated 
sky maps were generated using a \texttt{HEALPix} resolution of 
$N_\text{side}=256$, which corresponds to a per-pixel frequency-independent 
resolution of $\theta_{\text{\rm pix}}\approx 14'$.

\subsection{Foregrounds}\label{sec:foregrounds}
21cm foregrounds currently lack a detailed analytic description, making 
them difficult to separate from the cosmological HI signal. However, 
descriptive numerical simulations exist \citep[e.g.][]{hammurabi-gal-fg}, largely extrapolated from 
observed radio maps, such as the Haslam and Planck maps \citep{haslam1982, planck2016}. The foregrounds that we hope to remove from the 
observed sky can be separated into galactic and extra-galactic components. 
Extragalactic foreground sources are expected to be distributed according 
to a clear power spectrum \citep{Di_Matteo_2002, cohen_2004, santos-fg}, while galactic foregrounds, such as synchrotron 
emission, are expected to be localized, particularly in the galactic plane
\citep{santos-fg, hammurabi-gal-fg}.  

For galactic sources, the \texttt{CRIME} simulations extrapolate foregrounds 
from the 408 MHz map of \citet{haslam1982} to the relevant frequencies, as 
described in \citet{santos-fg}. For weaker foregrounds such as 
point sources and free-free emission, as well as for synchrotron effects on 
small scales, we adopt Gaussian realizations of the generic power-spectrum 
based model
\begin{equation}\label{eq:fg-cl-model}
    \begin{split}
    C_\ell (\nu_1, \nu_2) & ={}  A \left(\frac{\ell_{\rm ref}}{\ell}\right)^{\beta} \left( \frac{\nu^2_{\rm ref}}{\nu_1 \nu_2} \right)^\alpha \exp{\left(\frac{-\log^2(\nu_1/\nu_2)}{2 \xi^2} \right)}
    \end{split}
\end{equation}
with values given in Table \ref{tab:iso-params} (see \cite{santos-fg} for 
details). While we train our network on these fiducial values, we explore 
model generalization to new foreground parameters in 
Section \ref{sec:generalization}.

\begin{table}[t!]
\begin{center}
 \begin{tabular}{l c c c c}
 \toprule
 {Foreground Component} & $A\ \rm [mK^2]$ & $\beta$ & $\alpha$ & $\xi$ \\
 \midrule
Galactic Synchrotron & 1100 &  3.3 & 2.80 & 4.0 \\
 Point Sources & 57 & 1.1 & 2.07 & 1.0 \\
  Galactic free-free & 0.088 & 3.0 & 2.15 & 35 \\
    Extragalactic free-free & 0.014 & 1.0 & 2.10 & 35 \\
 \toprule
\end{tabular}
\caption{Fiducial foreground $C_\ell(\nu_1, \nu_2)$ model parameters used in this study, adapted from \cite{santos-fg} for the pivot values $\ell_{\rm ref} = 1000$ and 
reference frequency $\nu_{\rm ref}=130$ MHz.}\label{tab:iso-params}
\end{center}
\end{table}

\subsubsection{Polarized Foregrounds.}
Foreground polarization arises when synchrotron emitting electrons traverse 
the Milky Way's magnetic fields, changing their polarization angles due to
Faraday rotation \citep[see e.g.][]{RybickLightman86}. Despite some empirical observations
\citep[e.g.][]{wolleben-polar2006A&A...448..411W, deBruyn2006, schnitzeler2009}, this effect on radio foregrounds 
is poorly understood, except perhaps at very low radio frequencies, as purported by the Experiment to Detect the Global Epoch of Reionization Signature (EDGES) \citep{edges-2018}. Several models have been proposed to describe 
this phenomenon. The \texttt{CRIME} simulation package defines the Faraday 
depth at a distance $s$ along a line of sight (LOS), $\nhat$, as:
\begin{equation}\label{eq:faraday-depth}
    \psi(s, \nhat) = \frac{e^3}{2\pi (m_e c^2)} \int_0^s ds' n_e(s', \nhat)B_\parallel(s', \nhat)
\end{equation}
where $m_e$ is the electron mass, and $n_e(s, \nhat)$ $B_\parallel$ are the 
number density of electrons and galactic magnetic field contribution for the 
given LOS. As shown in \cite{Alonso_sim_2014}, the correlation over frequency 
for the polarization leakage field $\mu$ can then be written as:
\begin{equation}
    \langle \mu_{lm}(\psi) \mu^*_{l'm'}\rangle \propto \delta_{ll'} \delta_{mm'} \left( \frac{\ell_{\rm ref}}{\ell} \right) e^{-\frac{1}{2}\left[\frac{\psi - \psi'}{\xi_{\rm polar}}\right]^2}
\end{equation}
where the correlation length, $\xi_{\rm polar}$, and amplitude are free 
parameters.  The rest of the numerical values are given in 
\cite{Alonso_sim_2014}. The polarized emission correlation length is usually phenomenologically chosen. For example,  \cite{polar-Shaw_2015} use an equivalent correlation length scale $\xi_{\rm polar}$ of 
$0.1-0.05\ \rm rad\ m^{-2}$ to fit observations, while \cite{Alonso_sim_2014} choose a typical value of $0.5\ \rm rad\ m^{-2}$ to match simulations obtained by the \texttt{Hammurabi} code \citep{hammurabi-gal-fg}. We vary this parameter to probe failure modes 
in our foreground subtraction method in Section \ref{sec:polar}.

\subsection{Observational Noise}
The third component of the observed 21cm signal is (largely thermal) 
observational noise. Radio observational noise can be simply modeled 
as zero-centered Gaussian noise for single-dish experiments
\citep{bull-late-stage-cosmo-21, LiuShaw20}.


We modify the white noise model with a stochastic component in order to train and test cleaning methods on a wide range of possible observational thermal noise, capturing a range of possible current and future intensity mapping configurations. For each full-sky simulation, we adopt a frequency-dependent hierarchical noise model, which has the added advantage of better training our networks (see Section \ref{sec:deep21}):
\begin{subequations}\label{eq:noise-model}
\begin{align}
    \alpha_{\rm noise} &\curvearrowleft \log \mathcal{U}(0.05, 0.5) \\
    \sigma_{\rm noise} &= \alpha_{\rm noise}\ \langle T_b(\nu) \rangle \\
    \epsilon_{b,i} &\curvearrowleft\mathcal{N}(0, \sigma_{\rm noise} ) \\
    \hat{T}_{b,i} &= T_{b,i} + \epsilon_{b,i} \ \ \ \
\end{align}
\end{subequations}
where we relate the variance of the noise to the average fiducial cosmological temperature at a given frequency, $\langle T_b(\nu) \rangle$. The observed signal at pixel $i$, written as $\hat{T}_{b,i}$ is then given by the true signal, $T_{b,i}$, with the addition of the Gaussian noise $\epsilon_{b,i}$. For comparison, the noise model employed by \cite{Alonso_pca_2014} and \cite{paco-2016} correspond to an amplitude of $0.025 < \alpha_{\rm noise} < 0.12$. By sampling a large and competitive range of amplitudes for realizations of the per-pixel Gaussian noise, we allow the network to learn despite a variable range of noise. 

It should be noted that, strictly speaking, our noise model allows for the observed signal $\hat{T}_{b,i}$ to be negative, 
which is unphysical, but not an uncommon observable in radio astronomy as a result of readout noise \citep[see e.g.][]{wilson2011techniques}. However, given the range of $\alpha_{\rm noise}$ 
values taken above, this is a very rare occurrence.



\section{Foreground Removal Methods}\label{sec:methods}
Since foreground contaminants are several orders of magnitude brighter than
cosmological signal, the biggest challenge for upcoming observational data 
analysis will be to separate the two signals \citep{Di_Matteo_2002, pca_oliveira2008}. In this section we review blind 
data preprocessing for foreground subtraction and introduce our improved 
method. 



\subsection{Blind Foreground Subtraction}\label{sec:blind-techniques}
Fortunately, foregrounds are expected to be spectrally smooth in frequency \citep{Tegmark_2000, tegmark-liu12, planck2016}, 
while the cosmological signal is expected to vary with frequency according 
to Equation \ref{eq:temp-from-density} \citep{Di_Matteo_2002, LiuShaw20}. 
Current separation techniques therefore rely on the statistical
distinctions between 21cm spectral components.

As we outlined in the last section, the observed 21cm signal can be modeled 
as the sum of three components: cosmological, noise, and foreground modes. 
Formally, we write:
\begin{equation}\label{eq:sum_components}
    \begin{split}
    T_\text{obs}(\nu, \nhat) = T_\fgd(\nu, \nhat) + T_\cosmo(\nu, \nhat) + T_\text{noise}(\nu)
    \end{split}
\end{equation}
where each component is described with respect to a given frequency, $\nu$, 
and line of sight direction, $\nhat$. In a system of discrete frequencies, 
we can write a linear system for each line of sight over frequency:
\begin{equation}
    \textbf{x} = \boldsymbol{\hat{A}} \cdot \textbf{s} + \boldsymbol{\mathcal{C}_0}
\end{equation}
where each $x_i = T_{\rm obs}(\nu_i, \nhat)$, and $A_{ik} = f_k(\nu_i)$ and $s_k = S_k(\nhat)$ are linearly separable basis functions and foreground sky components, respectively. The cosmological signal and thermal noise can then be packaged as $\boldsymbol{\mathcal{C}_0} = T_\cosmo(\nu, \nhat) + T_\text{noise}(\nu)$. We see that in these terms, foreground subtraction becomes a residual learning problem, such that we aim to reconstruct $ \boldsymbol{\mathcal{C}_0} = \textbf{x} - \boldsymbol{\hat{A}} \cdot \textbf{s}$ as accurately as possible.

\subsection{PCA Residual Analysis}
\label{section:pca}

Principal Component Analysis (PCA) makes use of the statistical properties 
of foreground signals by simultaneously fitting foreground sky maps, $s_k$, 
and foreground functions, $A_{ik}$ \citep{pca_oliveira2008, Alonso_pca_2014}. Intuitively, PCA can be thought 
of as fitting a multidimensional ellipsoid to a feature space, with 
orthogonal axes (eigenvectors) pointing in the directions of the largest 
variance. PCA is an orthogonal transformation which maps the data from a set 
of basis vectors in which the data is correlated, to a basis in which the 
data is linearly uncorrelated. Since foregrounds are expected to be smooth 
and highly correlated in frequency \citep{Di_Matteo_2002, PritchardLoeb12,
pca_oliveira2008}, removing the components of largest eigenvalue (see Section 
\ref{sec:polar} in this work and Figure 1 in \citet{Alonso_pca_2014}) is expected 
to preserve the cosmological signal on large angular scales relevant to 
cosmology. The method has been employed for foreground cleaning in both 
simulated and real 21cm data 
\citep{chang-pca-real, Switzer_2015_pca_real, Masui_2013}.

\begin{figure*}[tb]
    \centering
    \includegraphics[width=\textwidth]{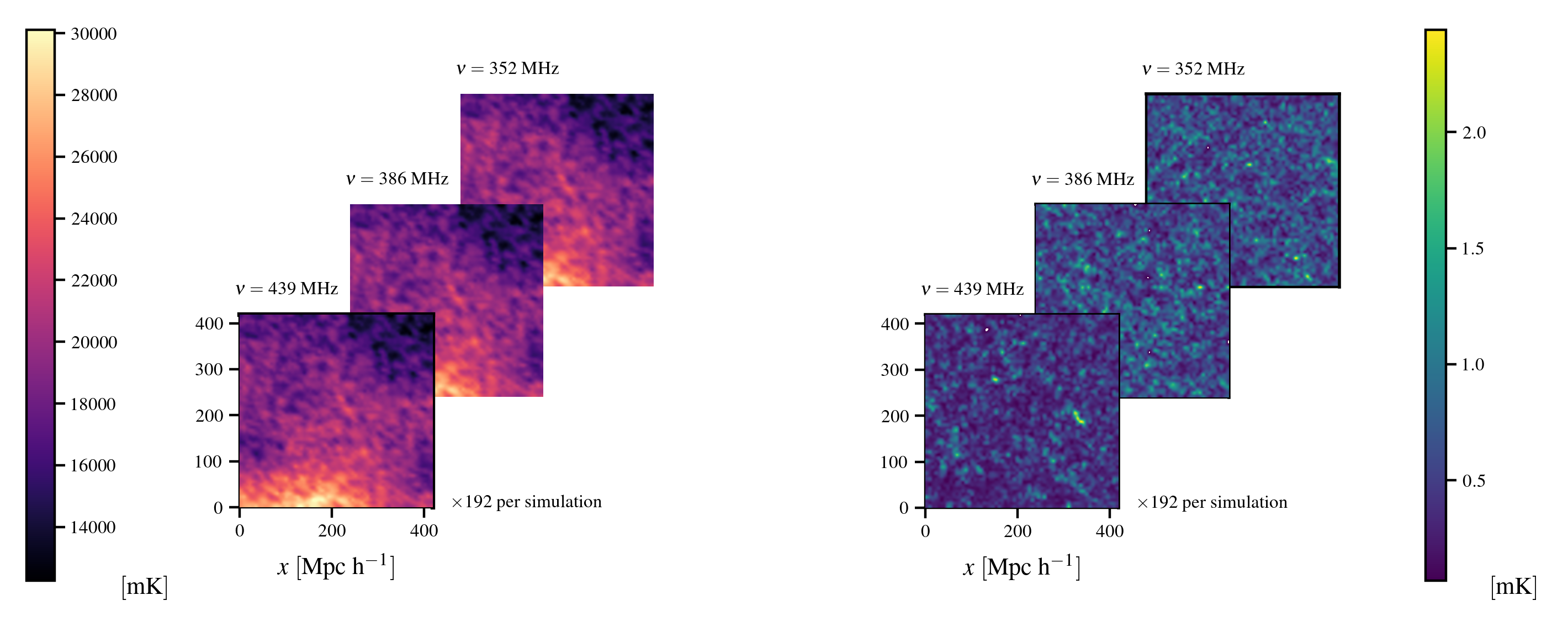}
    \caption{2D slices from input foreground (\textit{left}) and output cosmological (\textit{right}) voxels for the \deeep{} network. Each full-sky simulation is comprised of 192 \texttt{HEALPix} pixels at 690 different frequencies. We first diagonalize each sky in frequency and remove the first 3 principal components. We then take 64 frequencies from the first bin in Table \ref{tab:freq-bins} to generate 3D voxels of dimension $64^3$ for \deeep{} to process in batches of 16. Each epoch we process $80\times 192$ training and $10\times 192$ validation voxels.}
    \label{fig:inputs}
\end{figure*}

For our analysis, we repeat 
\citet{Alonso_pca_2014}'s PCA removal procedure here. First we bin our 
observed maps (foreground and cosmological signal) into $N_{\nu} = 64$ 
frequency bands. We define a correlation matrix, $\textbf{C}$ in frequency 
for all $N_{\rm pix}$ pixels in our simulation:
\begin{equation}\label{eq:pca-covariance}
C_{ij} = \frac{1}{N_{\rm pix}} \sum_{n=1}^{N_{\rm pix}}\ \frac{T(\nu_i, \hat{\textbf{n}}_n)\  T(\nu_j, \hat{\textbf{n}}_n)}{\sigma_i \sigma_j},
\end{equation}
where $T(\nu_i, \hat{\textbf{n}}_i)$ is the observed 21cm map signal and 
$\sigma_i$ are root-mean-square fluctuations of  $\boldsymbol{\mathcal{C}_0}$ 
in mK, in the $i$th frequency band. Each $\sigma_i$ is estimated iteratively 
from the data \citep{Alonso_pca_2014, bishop_prob_pca}. The covariance 
$\textbf{C}$ can then be diagonalized via eigenvalue decomposition:
\begin{equation}
\boldsymbol{\Lambda} = \textbf{U}\textbf{C} \textbf{U}^T = \textbf{diag}(\lambda_1, \dots, \lambda_{N_\nu}),
\end{equation}
where $\boldsymbol{\Lambda}$ is the diagonal eigenvalue matrix for 
$\textbf{C}$, and $\textbf{U}$ is an orthogonal matrix comprised of 
the corresponding eigenvectors. $\boldsymbol{\Lambda}$ is ordered by 
decreasing eigenvalues (principal components). For every pixel $n$ we 
compute the PCA spectrum in frequency, and then project onto 
$\boldsymbol{\Lambda}$. We then remove the first $N_{\rm comp}$ 
components, and generate a filtered spectrum from the remainder, 
$\boldsymbol{\mathcal{C}_0}$, which is then assigned to the pixel as 
the output.

For our analyses, we preprocess observed radio sky maps and remove both
the first three and first six principal components from our foreground 
maps. Henceforth, the notation PCA$-N_{\rm comp}$ corresponds to signal 
for which the first $N_{\rm comp}$ principal components have been removed. 

Despite its ability to recover cosmological statistics in some frequency 
ranges, as demonstrated in \cite{Alonso_pca_2014, pca_oliveira2008}, it 
is important to realize that blind subtraction techniques do not guarantee that HI overdensity anisotropy will be preserved at all scales, since they are not based on a signal likelihood. {PCA is best suited to removal of bright
foregrounds that occupy smoother, low-rank modes compared with the cosmological signal.} Unfortunately, the
cosmological signal
exhibits similar, smoothly varying structure on large scales, meaning PCA will remove 
cosmological clustering information needed for different studies (e.g. 
primordial non-Gaussianities \citep{gaussSekiguchi_2019}). Furthermore, 
polarization leakage from galactic synchrotron emission can create choppy 
foreground signal that is difficult for blind methods to distinguish from 
the cosmological signal (see \cite{villaescusanavarro2014crosscorrelating} 
and Section \ref{sec:polar}). This motivates finding an approach in which a separation
scheme is informed of the signal and foreground patterning.

\subsection{\deeep{} Neural Network}
\label{sec:deep21}
In this section we present our novel method for cleaning foregrounds from 
21cm maps. We adopt a convolutional neural network (CNN) architecture based on the UNet model of 
\cite{unet_review}, which maps images to images via a symmetric 
encoder-decoder convolution scheme. {This simulation-trained learning approach seeks to make a more informed separation of signals since the network has access to both foreground and cosmological axes in training.}

\subsubsection{Input Preprocessing}
\begin{figure*}[tp]
\includegraphics[width=\textwidth]{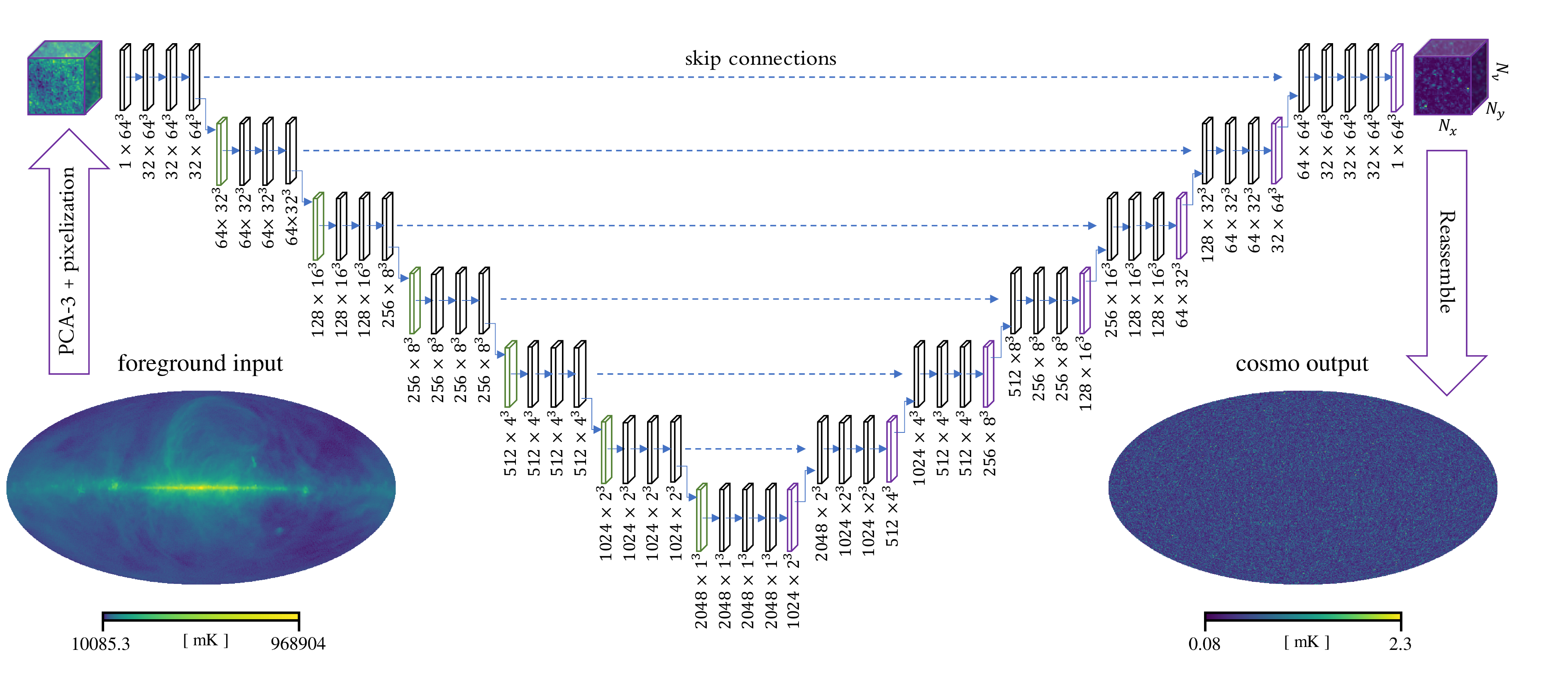}
\caption{UNet Architecture and training scheme. We first remove the first three principal components from the simulated observed maps. We then split each map via the \texttt{HEALPix} pixelization scheme for the network to process. On the encoder side, input data undergo \texttt{w}=3 convolutions (\textit{black prism}) at each level and are subsequently downsampled (\textit{green prism}) \texttt{h}=6 times, halving the spatial dimensionality while simultaneously doubling the number of filters at each level. The data are then decoded via symmetric transposed convolutions (\textit{purple prism}). Skip connections
concatenate features at each depth, allowing the network to learn a specific correlation scale at a time.}
\label{fig:experiment}
\end{figure*}

Since PCA has been shown to effectively remove the majority of foregrounds \citep{Alonso_pca_2014, pca_oliveira2008}, we focus our analysis on recovering the physical, cosmological signal from the PCA residuals. We feed in PCA-3 residual maps, processing input maps using \texttt{scikit-learn} \citep{scikit-learn}, which implements \cite{bishop_prob_pca}'s probabilistic PCA algorithm. Removing the first three components centers input signal on zero and drastically reduces input amplitudes, but does not remove too much small-scale cosmological clustering, as shown in \cite{Alonso_pca_2014}. PCA preprocessing thus has the added benefit of scaling inputs appropriately for neural networks, which perform best for inputs in the range $[-1, 1]$ \citep{GoodBengCour16}.
Unlike previous analyses (see e.g. \cite{Alonso_pca_2014, paco-2016, Cunnington_2019, pca_oliveira2008}), we do not perform instrument-dependent Gaussian beam smoothing as a preprocessing step, since we want to test the ability of our model to recover signal in the limit of pixel-size resolution.

\subsubsection{Dataset Assembly}
To test our foreground separation methods, we generated a suite of 100 full-sky cosmological and foreground \texttt{CRIME} simulations over 690 frequencies, $350 < \nu < 1044$ MHz, each separated by $\Delta \nu \sim 1 \rm MHz$. We then add a copy of the foreground and cosmological maps together, and split the dataset into 80 training simulations, 10 validation, and 10 (hidden) test simulations.

The UNet architecture is ideally suited to image-like data. For this reason, we split each simulation into 192 equal-area windows via the \texttt{HEALPix} pixelization scheme \citep{healpix}. We then stack maps in frequency, drawing 64 frequencies evenly within the designated redshift bin, yielding 192 cubic voxels of dimension $(N_{\theta_x}, N_{\theta_y}, N_\nu) = (64, 64, 64)$ pixels, shown schematically in Figure \ref{fig:inputs}. According to the train-validation-test split, the network sees $80\times192=15,360$ training voxels each epoch, followed by $10\times192=1,920$ validation voxels. We set aside 10 hidden test simulations with which to assess our cleaning methods.


The UNet architecture maps input voxels to output voxels via a contracting path, in which input dimensionality is halved and feature channels doubled iteratively via stride-2 downsampling convolutional layers, depicted schematically by green prisms in Figure \ref{fig:experiment}. At each depth of the network, we perform \texttt{w} convolutions. Data are then upsampled via transposed convolutions through a symmetric upsampling path. Skip connections concatenate features from one side of the network to the other, allowing each depth of the network to focus on learning a specific scale of correlations at a time. These correlations are then summed as the network upsamples on the output side. We employ batch normalization and \texttt{ReLU} activation between convolutional layers, except for the last convolutional block on the output side. Once voxels have been processed by the network, we reconstruct the full-sky cosmological maps to compute power spectra and clustering statistics. 

\subsubsection{Loss Function}
To train the networks, we would like to minimize a pixel-wise loss function of the form $\mathcal{L}(p,t) = \sum_i L(|p_i - t_i|)$ between prediction, $p$, and simulation target, $t$ of each $i^{th}$ voxel, and $L(x)$ is the pixel-wise loss function. We find empirically that the standard Mean Square Error (MSE) loss proved volatile early in training. For this reason we selected the Log-Cosh loss function 
\begin{equation}\label{eq:lossfn}
\mathcal{L}(p,t) = \sum_i \log \cosh (p_i - t_i)\, .
\end{equation}
This function behaves much like the L1 norm for poor predictions (large values of $|p_i - t_i|$), making it robust to outliers, and approaches $(p_i - t_i)^2 / 2$ for small residuals. For network performance validation and test statistics we look at the Log-Cosh loss, as well as the standard MSE metric.

\subsection{Training Procedure}
In training our selected architecture we make use of a combination of the \texttt{AdamW} optimizer \citep{adamw} and a step-wise learning rate reduction conditioned on validation data. This choice of learning routine provides weight decay regularization, as well as a fine-tuning of network optima. 

Every training epoch the network sees 80 simulations of foregrounds added to cosmological signal, subject to a new observational noise realization for a sampled $\alpha_{\rm noise}$. The observed maps are first preprocessed by the PCA-3 subtraction and then pixelized into \texttt{HEALPix} voxels described above. The PCA-3 residuals are then processed by the UNet network.

We split our frequency range to test our foreground cleaning in the context of analyses such as \cite{paco-2016}. Redshift shells and corresponding co-moving distances are shown in Table \ref{tab:freq-bins}. For our analysis we focus on evaluating network performances in the co-moving shell of lowest frequency, since these high-redshift regions are interesting for both Baryonic Acoustic Oscillation (BAO) measurement, as well as post-EoR structure analysis \citep{PritchardLoeb12}. Furthermore, these regions have consistently proven difficult for blind foreground techniques to clean, especially in angular power spectrum recovery \citep{Alonso_sim_2014, paco-2016}. 

\subsection{Hyperparameter Tuning}
In choosing our network architecture, we first compared architectures with 2D and 3D convolutional kernels with different network depths. We anticipate that training 3D convolutional kernels perform better than 2D convolutions, since inputs in this scenario are treated as full 3D volumes, capturing frequency patterning in $\nu$, as well as angular patterns in $\theta_x$ and $\theta_y$.

Other important hyperparameters we considered were UNet depth, \texttt{h}, or number of down-convolutions (denoted by green prisms in Figure \ref{fig:experiment}), and the number of convolutions at a given dimension, \texttt{w} (convolutional block width). To test hyperparameters, we developed a dynamic UNet model compatible with the \texttt{HyperOpt} Python library \citep{hyperopt}. Our architecture draws hyperparameters from proposal distributions and trains the resulting architecture on a smaller set of training data. We selected the hyperparameter combination that yielded the lowest validation loss after testing 550 trial architectures. The priors from which we drew our hyperparameters are listed in Appendix \ref{sec:appendix}. The results yielded the interesting result that deeper, wider UNets outperformed shallower networks (see \cite{network-depth-mhaskar2016deep} for a theoretical investigation). 
We found that in particular, architectures with network width $2 < \texttt{w} < 4$ and height satisfying 
\begin{equation}\label{eq:height}
    \texttt{h} = \log_{\rm \texttt{stride}} \texttt{n}_{\rm filters}; \ \ \ \rm \texttt{stride} = 2
\end{equation} consistently yielded the lowest losses. Too many convolutions per block frequently obscured the sharp $T_b$ distribution (see Figure \ref{fig:temp_comp}), and Equation \ref{eq:height} guarantees an architecture that compresses inputs down to dimension $1^3$ for stride-2 down-convolutions, meaning the network learned correlations on a pixel-sized scales. Our optimized architecture, henceforth \texttt{deep21}, is displayed graphically in Figure \ref{fig:experiment}.

Neural network outputs are generally not probabilistically interpretable \citep{charnock2020bayesian}, and no tractable image-producing Bayesian neural networks currently exist,
so quantifying the variability of foreground cleaning on a given dataset is not possible with a single UNet. 

Motivated by the methodology of deep ensembles \citep{deep_ens, deep_ens_loss_land} we train an ensemble of $M=9$ networks with independently Glorot-Uniform-initialized weights \citep{Glorot10understandingthe} for 300 epochs in parallel.
The \texttt{HEALPix} data inputs are also subject to the stochastic noise model as before, as well as random sky-sized rotations on the sphere, such that the networks learn to denoise pixels independently of orientation.  
To gauge the uncertainty of the cleaning method, each ensemble member then performs foreground separation on ten test simulations, subject to competitive observational noise with a fixed $\alpha_{\rm noise} = 0.25$, falling in the upper range of our amplitude prior and roughly twice the maximum noise amplitude utilized by \citet{Alonso_pca_2014}. This practice allows us to estimate the epistemic uncertainty on the summary statistics obtained from the predicted full-sky maps.

\section{Results and Analysis}\label{sec:results}

\begin{figure*}[htpb]
    \centering
    \includegraphics[width=\textwidth]{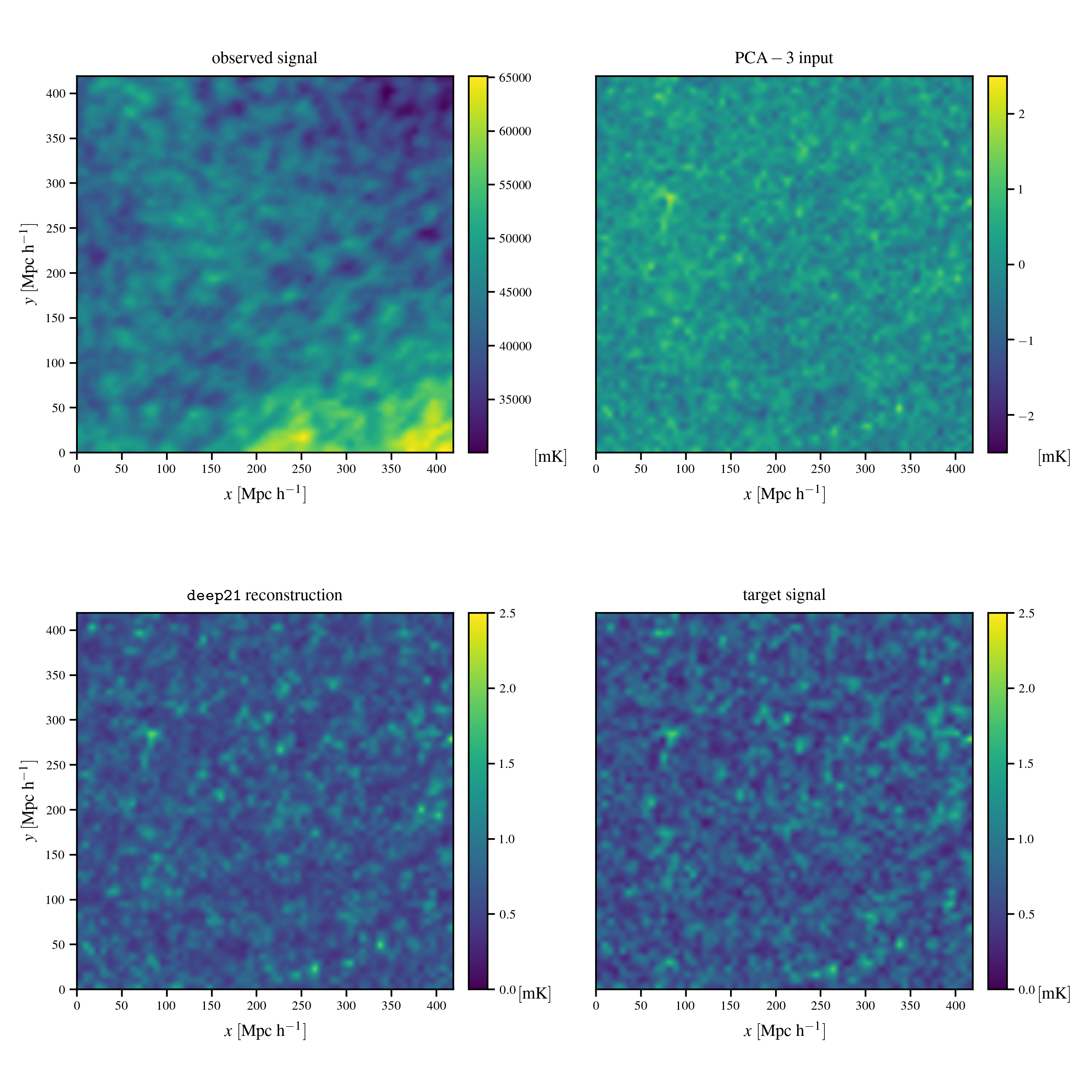}
    \caption{
    {2D slices  at $\nu=392\rm\  MHz$ comparing raw foreground signal (\textit{top left}), PCA-3 UNet inputs
    (\textit{top right}), to the UNet ensemble prediction (\textit{lower left}) and target cosmological signal
    (\textit{lower right}). \deeep{} is able to correctly reconstruct the signal from minimal PCA-3 subtracted
    inputs. }
    }
    \label{fig:slice-comp}
\end{figure*}

\subsection{Visual Inspection}
\label{section:visual}
Most current and upcoming cosmological experiments are aimed at reconstructing relevant clustering statistics from brightness temperature maps. Therefore it is prudent to ensure that our network outputs clean maps that capture temperature distributions at given frequencies. An initial check for \texttt{deep21}'s performance is a qualitative one: Figure \ref{fig:slice-comp} shows each cleaning method's performance on a \texttt{HEALPix} pixel slice drawn from the test set. The top two panels display observed input signal and the initial PCA-3 preprocessing network inputs. The bottom row compares the network-cleaned panel compared with the target cosmological signal. \deeep{} is able to recover intricate cosmological signal from corrupted inputs almost indistinguishable from the simulated targets.

\begin{figure*}[htpb]
    \centering
    \includegraphics[width=\textwidth]{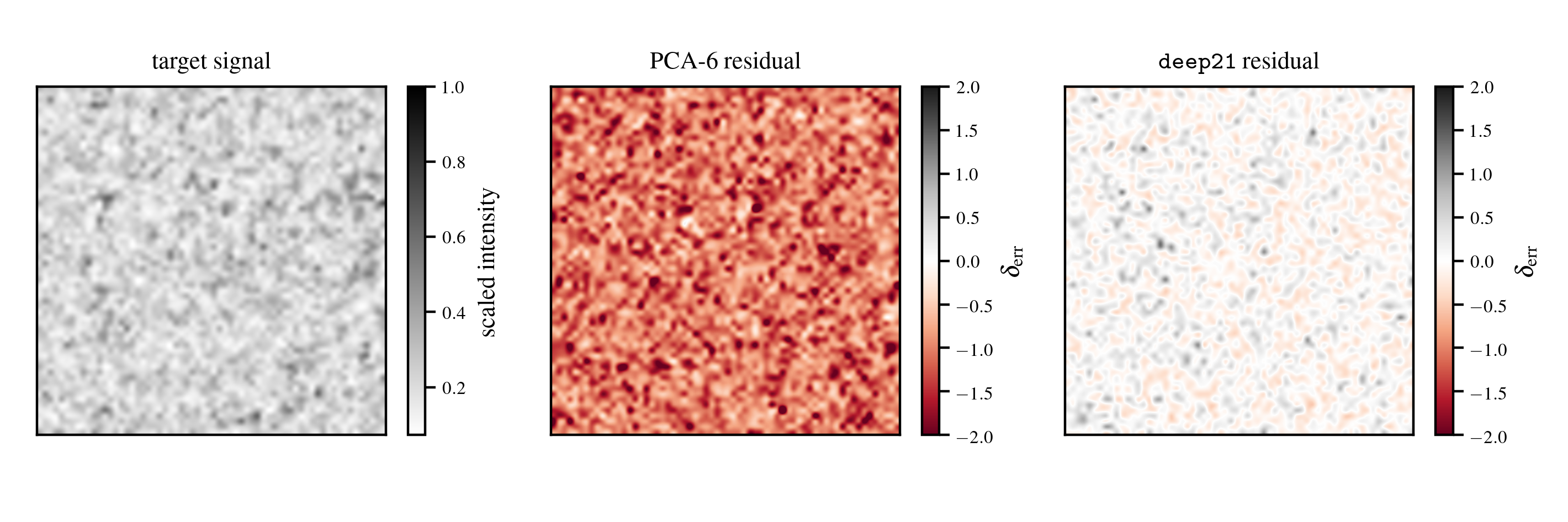}
    \caption{Temperature map residuals for PCA-6 (\textit{middle}) and \deeep{} (\textit{right}) map residuals compared to scaled intensity of the simulated cosmological signal (\textit{left}) for the same 2D slices as Figure \ref{fig:slice-comp}. The UNet ensemble recovers a much more accurate tomography than the PCA method, the latter failing to capture details around high-intensity regions and voids. Removal of the first moment in the PCA method results in a significant deviation ($\sim 200 \%$) from the true signal, while \deeep{} predictions yield small, positive residuals.
    }
    \label{fig:res-slice-comp}
\end{figure*}

In Figure \ref{fig:res-slice-comp}, we compare the per-pixel scaled cosmological signal (greyscale panel) to the relative residual error for cosmological brightness temperature $T_b$,
\begin{equation}
    \delta_{\text{err}, i} = \frac{p_{i} - t_i }{ {t}_{i}}
\end{equation} where $p_i$ and $t_i$ are the pixel-wise predicted and target signals, respectively. Here we compare the best-case (PCA-6) blind subtraction to \deeep{}. We note that PCA  predictions over-subtract the signal, particularly in low-density regions. \deeep{}'s residuals are all well within order unity of the target, with some deviations above zero.

\subsection{Clustering Statistics}
\label{sec:clustering-stats}

The most important cosmological parameter constraints from HI intensity mapping will most likely come from power spectra of the 21cm brightness temperature, since two-point correlation functions contain the vast majority of information regarding underlying cosmology on large, linear, scales. For this study, we consider angular and radial power spectra separately, capturing clustering patterning on the sky and along each line of sight, respectively.

For a fixed frequency and assuming a full-sky survey, the angular power spectrum of the brightness fluctuations $\Delta T_b$ is computed first by calculating the spherical harmonic components:
\begin{equation}
    a_{\ell m}(\nu) = \int d \nhat^2 \Delta T_b(\nu, \nhat) Y^*_{\ell m}(\nhat),
\end{equation}
where $Y_{lm}(\nhat)$ are the spherical harmonic basis functions. We can then estimate the power spectrum by averaging over the moduli of the harmonics:
\begin{equation}\label{eq:Cl-def}
    \widetilde{C}_l = \frac{1}{2\ell + 1} \sum_{m=-\ell}^\ell |a_{lm}|^2
\end{equation}
where small $\ell$ correspond to the largest scales. We calculated the angular power spectra for our maps using the \texttt{healpy} Python library \citep{healpy}.

To capture radial clustering in an HI survey independent of redshift effects, one must make two assumptions, namely 1) each line-of-sight (\texttt{HEALPix} pixel) window satisfies the flat-sky assumption \citep{Alonso_sim_2014}, and 2) that the redshift bin under consideration is narrow enough that no significant cosmological expansion occurs between the edges of the bin. The resulting power spectrum then describes the clustering distribution independently of cosmological expansion effects. Given these two assumptions, we can average over all possible radial lines of sight, $i = 1, \dots, N_\theta$, to obtain the radial power spectrum:
\begin{equation}\label{eq:radial-def}
    P_\parallel(\kpara) = \frac{\Delta \chi}{2 \pi N_\theta} \sum^{N_\theta}_{i=1} | \widetilde{\Delta T_b}(\nhat, \kpara) |^2
\end{equation}
where the Fourier coefficients $\widetilde{\Delta T_b}$ are estimated using the Fast Fourier Transform (FFT) over each line of sight, and $\Delta \chi = \chi(z_\text{max}) - \chi(z_\text{min})$ is the comoving width of the given redshift bin. Given a constant frequency interval separating the spherical surfaces, $\delta \nu$, the same interval is expressed in the conjugate space as $\delta k_\nu = 2\pi / \Delta \nu$. The radial coordinate, $\kpara$, can then be defined as \citep{Alonso_sim_2014}:
\begin{equation}
    \kpara = \frac{\nu_{21} H(z_{\rm eff})}{(1 + z_{\rm eff})^2} k_\nu,
\end{equation}
where $z_{\rm eff}$ is the effective redshift for the comoving volume under consideration. 

\begin{table}[t!]
\centering
\adjustbox{width=0.55\textwidth}{%
 \begin{tabular}{c c c c c}
 \toprule
 {$\nu$ (MHz)} & {z} & ${\langle z \rangle}$ & Vol. $(h^{-1} \text{Gpc})^3$ & $N_\text{side}$ \\
 \hline
    $[886-1044]$ & $[0.36-0.60]$ &0.47 & 12 &  256 \\
    $[667-886]$ & $[0.60-1.12]$  & 0.87 & 50 & 256 \\
    $[476-667]$ & $[1.12-1.88]$  & 1.50 & 108 & 256 \\
    $[350-491]$ & $[1.88-3.05]$  & 2.47 & 187 & 256 \\
 \toprule
\end{tabular}
}
\caption{Comparison of the four redshift co-moving shells used in the UNet assessments. Co-moving shells were chosen by splicing the simulation into bins with equal numbers of frequency. Our analysis focuses on the highest-redshift bin because this is where blind foreground techniques such as PCA reduction have been shown to perform the worst in the literature.}
\label{tab:freq-bins}
\end{table}


To capture uncertainty over the space of the \deeep{} ensemble, we compute a weighted average, $\Bar{Z}_w$, and standard deviation, $\sigma_w(Z)$, of each network's independent estimate of a given statistic, $Z$. We employ proper scoring weights by computing the inverse globally-averaged MSE for each network's prediction over test data: $w_{\rm m} = \frac{1}{\langle {\rm MSE}\rangle}$ for $m=1,\dots 9$ independent networks. Here we do not explicitly assume a Gaussian form, so $\sigma$ does not correspond to the 68\% inclusion interval. We note that to apply this procedure in a real observational setting, scores for each ensemble member would need to be computed for a set of validation simulations, and then used to compute statistics obtained from the cleaned observed sky.

\begin{figure*}[htp]
    \centering
    \includegraphics[width=\textwidth]{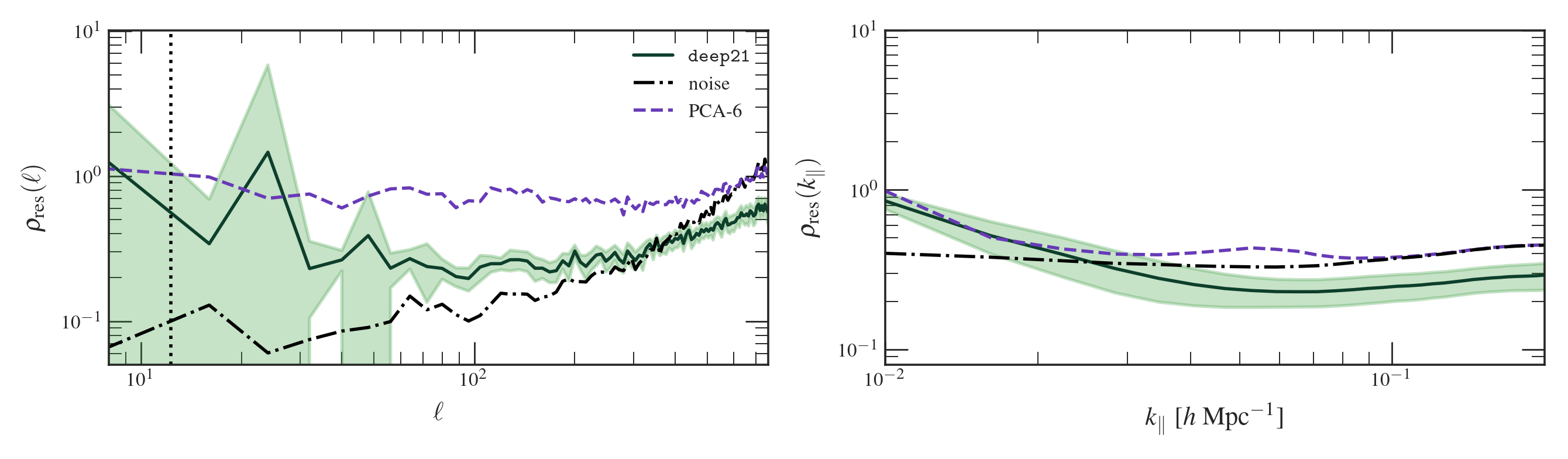}
    \caption{Comparison of angular (\textit{left}) and radial (\textit{right}) residual map power spectra as a fraction of the target cosmological signal predicted by the \texttt{deep21} UNet ensemble (\textit{green}) blind PCA reduction (purple), and noise realization (\textit{black}) with $\alpha_{\rm noise} = 0.25$ over a single full-sky test simulation. Confidence intervals corresponding to $\pm 2 \sigma_w$ over the space of ensemble parameters is estimated in shaded green. We display \deeep{}'s angular resolution via the black dashed vertical line.
    The angular power spectrum shown is computed for a single frequency, $\nu = 357\ \rm MHz$, while the radial power spectrum is computed for the lowest frequency bin, with mean redshift $\langle z \rangle = 2.5$. The network successfully learns to marginalize out additive observational noise in the radial direction at smaller scales ($k_\parallel > 0.015$).}
    \label{fig:res-power-spec}
\end{figure*}

We consider the power spectra calculated for the residual maps for each cleaning method, defined for $P \in \{C_\ell, P_\parallel \}$ as:
\begin{equation}
    \rho_{\rm res} = \frac{P_{\rm res}}{P_{\rm cosmo}} = \frac{P(p - t)}{P(t)}
\end{equation}
where $t$ is the target cosmological signal and $p \in \{ T_{\rm PCA}, T_{\rm \deeep{}}\}$ is the given cleaning method's predicted map. We additionally consider $\rho_{\rm res}$ computed for the noise map, $P(T_{\rm noise})$, generated for the test simulation.

This statistic quantifies each cleaning method's residuals as a fraction of the true cosmological signal in both the angular and radial directions. The noise residual demonstrates to what degree observational error obscures the structure estimate.

We compare \deeep{} to the PCA-6 residual noise realization generated with $\alpha_{\rm noise} = 0.25$ in Figure \ref{fig:res-power-spec}. \deeep{} (green) outperforms PCA (purple) in both the angular and radial directions, and successfully fits cosmological signal at small scales despite high observational noise contribution (black dashed line). We interpret this result as a successful marginalization of the observational noise. Through training, \deeep{} has learned to distinguish cosmological clustering from noise fluctuations at small scales.
By contrast, the PCA-6 residual asymptotically approaches the noise boundary in both plots, indicating a limit to the blind foreground cleaning at small scales.
\deeep{} also substantially reduces the loss of signal at large radial scales incurred by the PCA method. The larger PCA-6 residual at small $k_\parallel$ indicates large-scale information lost to the foreground subtraction as demonstrated in \cite{Alonso_pca_2014}.


\subsection{Intensity Distributions}
\begin{figure*}[htb]
    \centering
    \includegraphics[width=\textwidth]{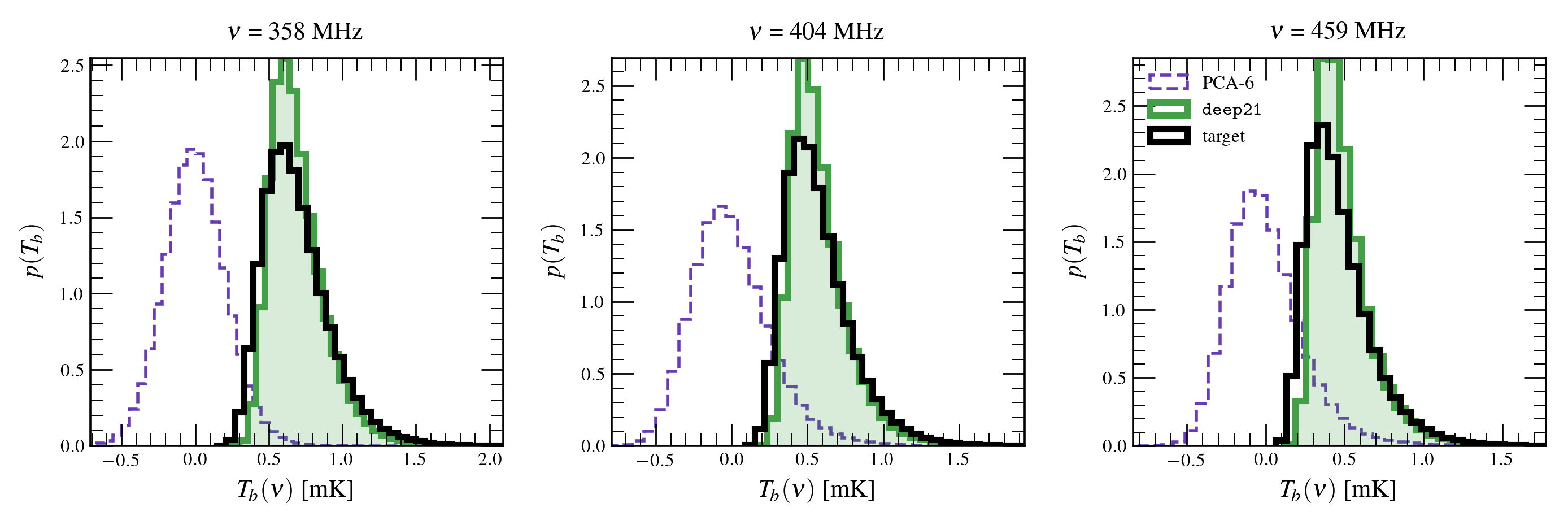}
    \caption{Comparison of the distribution of pixel temperatures from the PCA-6 (\textit{purple}) and \deeep{} (\textit{green}) cleaned maps to those of the cosmological simulations (\textit{black}) at several different frequencies. {\deeep{} captures the target asymmetric temperature PDF much more effectively than the PCA.}}
    \label{fig:temp_comp}
\end{figure*}

We also compare how well each cleaning method captures the distribution of the cosmological signal at each frequency. Figure \ref{fig:temp_comp} compares cosmological temperature distributions at several frequencies throughout a single test simulation. We see that the PCA method reproduces a more symmetric temperature distribution that is zero-centered. This result is anticipated by the definition of the method (which removes the mean of the distribution to diagonalize the signal in frequency). {It is clear from comparison to the true signal that \deeep{} captures the asymmetric distribution much more accurately than the PCA.}

\subsection{Power Spectrum Recovery}\label{section:clustering}

We additionally report clustering statistics based on the analysis by \cite{Alonso_pca_2014}. We introduce the power spectrum correlation statistic
{\begin{equation}\label{eq:epsilon}
    r(k) = 
    \frac{P_{\rm p \times t}(k)}{\sqrt{P_{\rm p}(k) P_{\rm t}(k)}} 
\end{equation} 
where $P_{\rm p}$ is the predicted auto-power spectrum from the foreground cleaning method under consideration, $P_{\rm t}$ is the auto-power spectrum generated by the target cosmological map, and  $P_{\rm p \times t}$ is the cross-spectrum between the predicted and target maps. This statistic quantifies discrepancies in phases introduced by each cleaning method. We also define the transfer function for coordinate $k \in \{ \ell, \kpara \}$ }
\begin{equation}\label{eq:transfer-fn}
    {\rm T}(k) = \sqrt{\frac{P_{p}(k)}{P_{\rm t}(k)}}
\end{equation}
which quantifies discrepancies in amplitude as a function of scale $k$ between cleaned and target maps. For a perfect foreground cleaning, both $r(k)$ and $T(k)$ approach 1.

\begin{figure*}
    \centering
    \includegraphics[width=\textwidth]{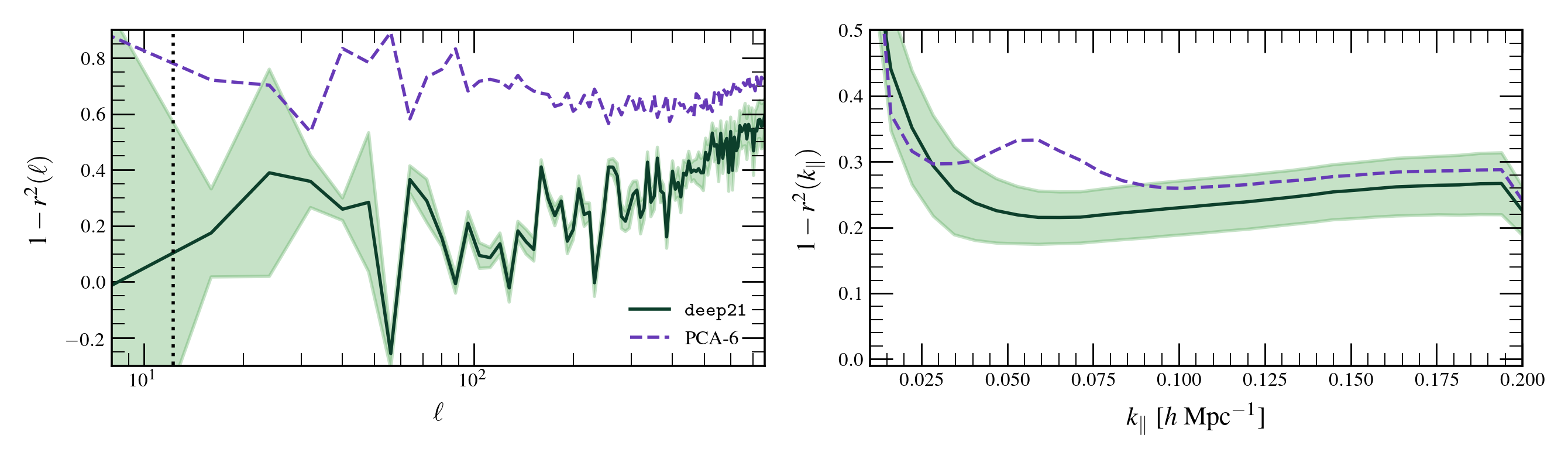}
    \caption{{Residual variance $1-r^2(k)$ for both angular (\textit{left}) and radial (\textit{right}) power spectra. The angular spectrum is computed at $\nu = 357\ \rm MHz$, with the angular resolution of a \deeep{} input voxel shown via the black dashed vertical line. \deeep{} (\textit{green}) outperforms the PCA-6 subtraction on all angular scales. In the radial direction, \deeep{} offers a significant improvement in preserving phase information on intermediate scales.}}
    \label{fig:power-spec-res}
\end{figure*}

{The statistic $1 - r^2(k)$ describes the fraction of variance in cleaned maps that isn't accounted for in the target map. We compare this variance as a function of scale for angular and radial power spectra in Figure \ref{fig:power-spec-res}. \deeep{} (green) outperforms the PCA-6 consistently at both large and small angular and radial scales, capturing the correct phase within $\sim 80\%$ accuracy within the comoving bin. This statistic demonstrates the results displayed in Figure \ref{fig:temp_comp} as a function of scale: network-based cleaning captures the non-Gaussian cosmological signal with much higher accuracy, particularly in the angular direction. This is because unlike the PCA, \deeep{}'s convolutional filters have access to neighboring pixels in the spatial axes, meaning foreground separation can be learned simultaneously in $\{\theta_x, \theta_y, \nu \}$}.

The consistency of the deep learning approach should be emphasized here: the \deeep{} method is largely scale-independent in foreground removal and signal reconstruction. Moreover, these results are not reliant on statistical marginalization over many data realizations like the analyses done in \cite{Alonso_pca_2014} and \cite{paco-2016}, meaning fewer observations need to be made in a realistic setting in order to achieve a consistent, successful foreground removal with estimated uncertainties. Furthermore, with more computational power, larger voxels can be expected to be processed in the future, resulting in an improvement in large-scale network recovery.

\subsection{Noise Performance and Instrument Effects}

\begin{figure}
    \centering
    \includegraphics[]{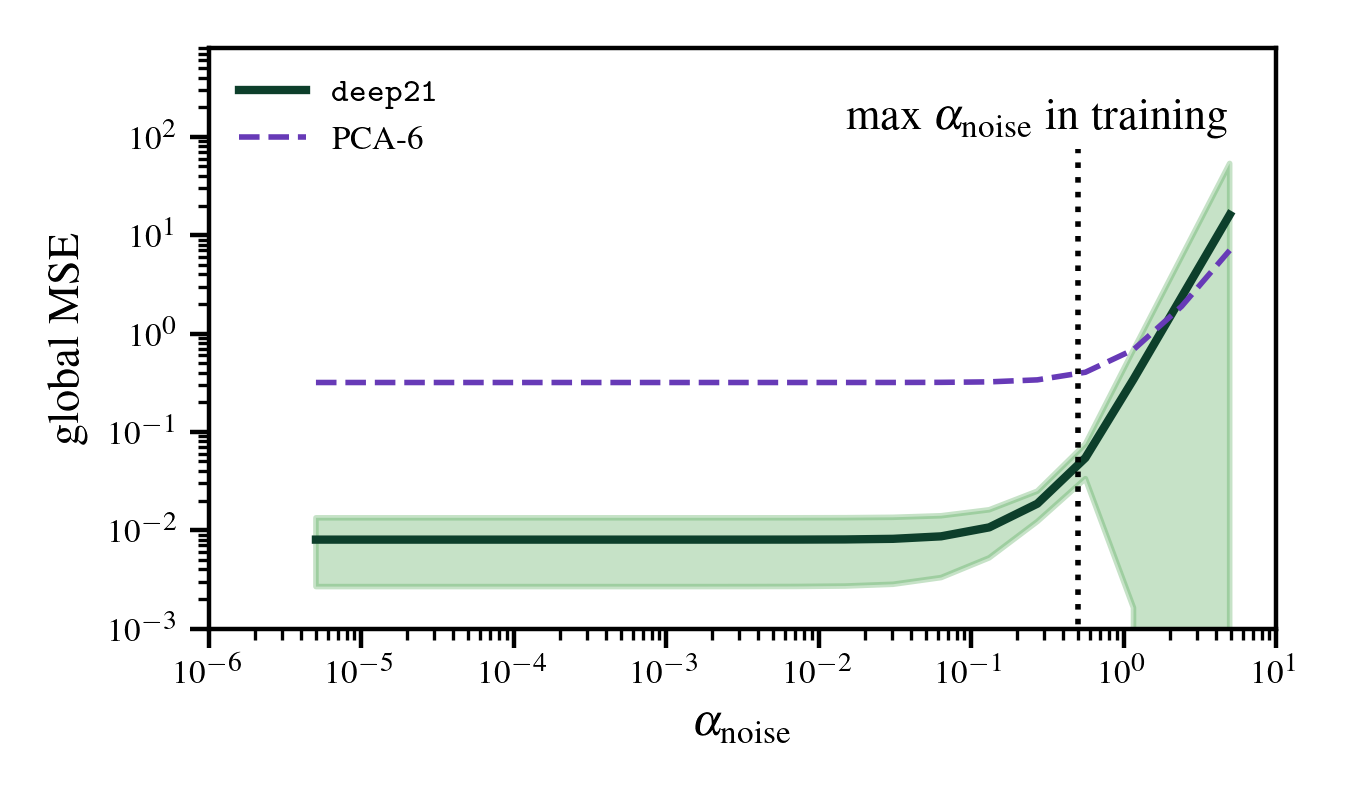}
    \caption{Noise performance testing for \deeep{} and PCA-6 cleaning routines. 20 maps were generated with increasing noise amplitude $\alpha_{\rm noise}$. As anticipated, \deeep{} performs consistently for $\alpha < \alpha_{\rm max} = 0.5$, since this represents the level of noise the network encountered during training according to Equation \ref{eq:noise-model}. The network's error quickly increases in variance and magnitude beyond this threshold. 
    }
    \label{fig:noise-test}
\end{figure}

\deeep{}'s training procedure includes variable levels of observational noise to improve test performance and incorporate uncertainty regarding noise levels in upcoming intensity mapping experiments. To test whether or not the ensemble learned to marginalize out this effect, we tasked the trained network with cleaning the same foreground simulation with a variable noise amplitude, $\alpha_{\rm noise}$. We plot the corresponding MSE metric in Figure \ref{fig:noise-test}. The network's test MSE remains fairly constant until we exceed the noise threshold, $\max\lbrace\alpha_{\rm noise}\rbrace = 0.5$, encountered in training. This shows that \deeep{} indeed captures the statistical properties of the observational noise it encountered during training, and subsequently marginalizes over it. In contrast, the blind PCA subtraction, whose MSE is dominated by the removal of the first moment of signal, does not increase in MSE until $\alpha_{\rm noise} = 1$, or order unity with the mean cosmological signal at a given frequency (see Equation \ref{eq:noise-model}). {This result shows that a network ensemble can be trained to be robust to observational noise and more complicated instrument effects, so long as a statistical model is specified and varied enough during training, a significant advantage over blind approaches with no access to this information.
}

\subsection{Generalization to new foreground parameters}
\label{sec:generalization}

\begin{figure*}[htpb]
    \begin{subfigure}{0.49\textwidth}
        \includegraphics[width=\textwidth]{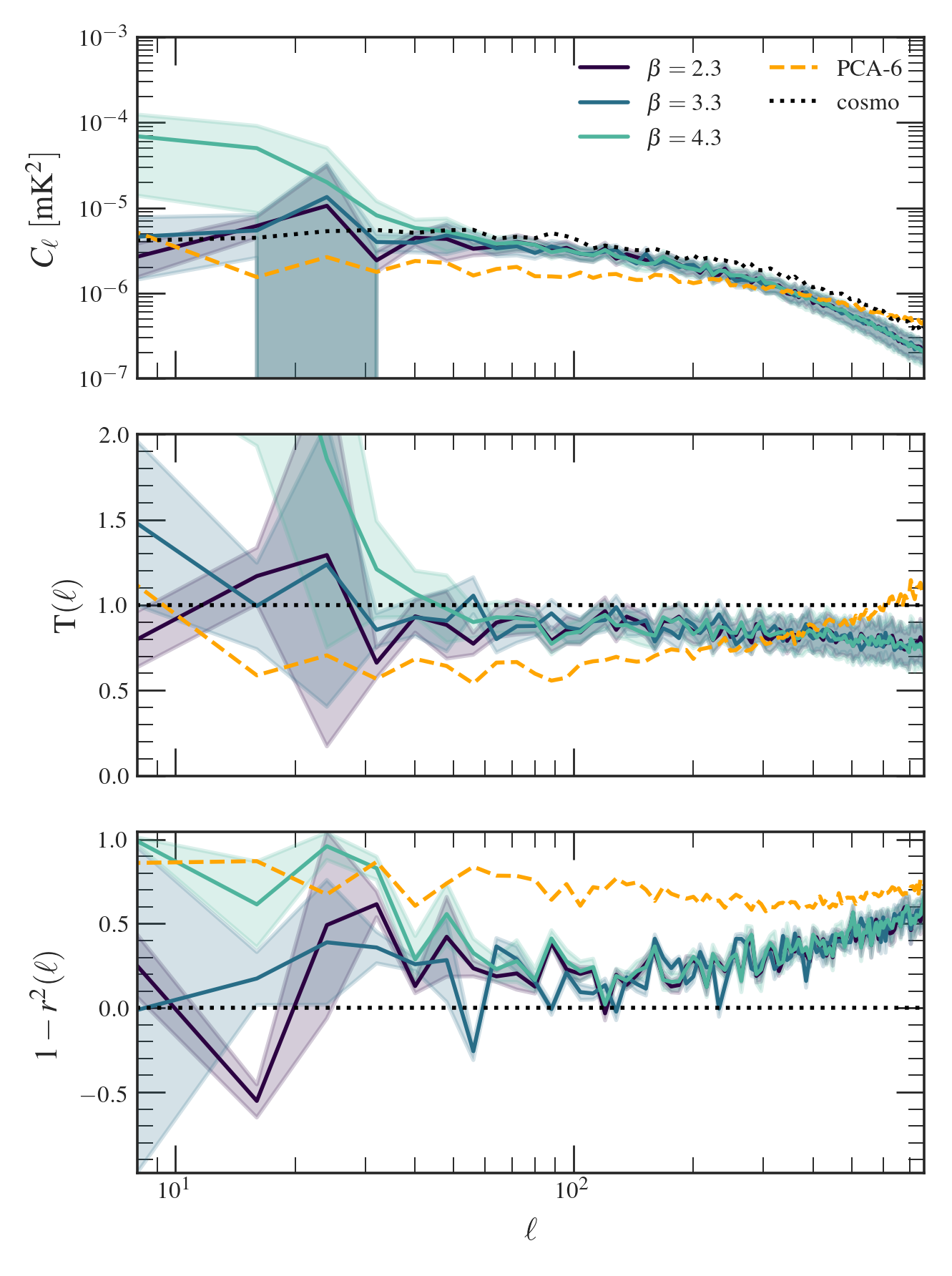} 
        \caption{Varying galactic synchrotron $\ell$ dependence.}
        \label{fig:beta}
    \end{subfigure}
    \begin{subfigure}{0.49\textwidth}
        \includegraphics[width=\textwidth]{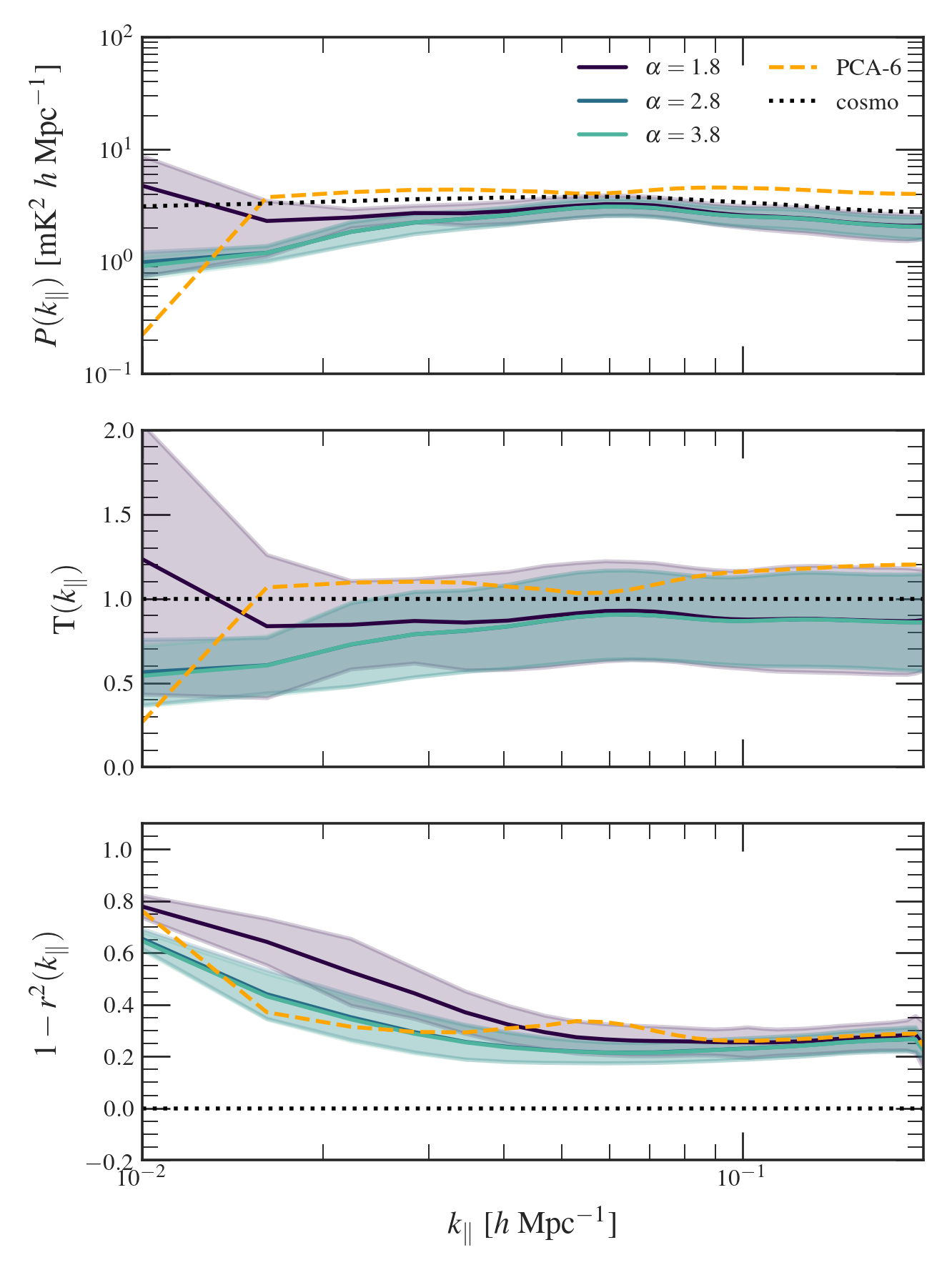}
        \caption{Varying galactic synchrotron $\nu$ dependence.}
        \label{fig:alpha}
    \end{subfigure}
    
    \caption{Generalization testing for \deeep{}. We compare two-point statistics on test data by varying foreground parameters, keeping training data and trained network ensembles constant. We vary the galactic synchrotron $\beta$ and $\alpha$ in Figures \ref{fig:beta} and \ref{fig:alpha}, respectively, according to Equation \ref{eq:fg-cl-model}. {We display power spectra (\textit{top}), transfer function, i.e. the square root of predicted to true spectrum (\textit{middle}), and residual variance fraction $1-r^2(k)$ (\textit{bottom}) for both angular (at $\nu=357\ \rm MHz$) and radial statistics.} 
    }
    \label{fig:generalization}
\end{figure*}

Despite training \deeep{} on many foreground and cosmological realizations, we assumed the same fiducial model for all training data generation. For a foreground cleaning experiment with real data, one would ideally train \deeep{} on a range of foreground and cosmological models. As a final test for our trained UNet ensemble, we task the network, trained on fiducial simulation parameters, with cleaning observed signal generated by different input parameters. 

To do this, we chose to alter the galactic synchrotron foregrounds via the \texttt{CRIME} simulation package, since these represent the largest of the foreground contaminants. Since the PCA processing removes synchrotron amplitude information, we elected to alter its correlation structure according to Equation \ref{eq:fg-cl-model}, namely 1) angular scale dependence, 2) frequency dependence and 3) switching on polarization effects. We display recovered power spectra for the correlation structure analysis in Figure \ref{fig:generalization}.

\subsubsection*{Varying  galactic synchrotron angular correlation dependence} Having trained \deeep{} on foregrounds with fiducial galactic synchrotron $\beta_o = 3.3$, we vary the synchrotron $\ell$ dependence according to Equation \ref{eq:fg-cl-model} by $\pm 30$\% to $\beta = 1.3 \beta_o$ and $\beta = 0.7 \beta_o $, generating a new full-sky simulation of 192 \texttt{HEALPix} voxels for each case. Increasing $\beta$ yields a smaller foreground correlation in $\ell$, reflected in \deeep{}'s over-estimate of the power spectrum in Figure \ref{fig:beta}. Likewise, the network under-estimates the power spectrum for lower $\beta = 2.3$.  This indicates that the network has indeed captured the fiducial foreground model in the training data.

\subsubsection*{Varying galactic synchrotron frequency correlation dependence} We repeated the same analysis for the $\alpha$ parameter, varying the galactic synchrotron $C_\ell$ model's dependence on frequency by $\pm 35\%$. We recovered a similar trend in \deeep{} performance in Figure \ref{fig:alpha}. Here, decreasing $\alpha$ results in a smaller correlation amplitude as a function of frequency. \deeep{} produces an over-estimate of the radial power spectrum since it is trained on $\alpha = 2.8$. Increasing $\alpha_v = 1.35 \alpha_o$ produces little difference in the ensemble estimate, indicating that the model might generalize well in this regime. 

In contrast, PCA-6 cleaning (dashed orange) yields almost identical results for changes in the correlation parameters, indicating that the blind method is robust to changes in correlation structure. We display foreground cleaning results at different scales for our test and generalization cases in Table \ref{tab:results}.

\subsubsection*{Polarized foregrounds}\label{sec:polar}
\begin{figure}[htpb]
    \centering
    \includegraphics[width=\textwidth]{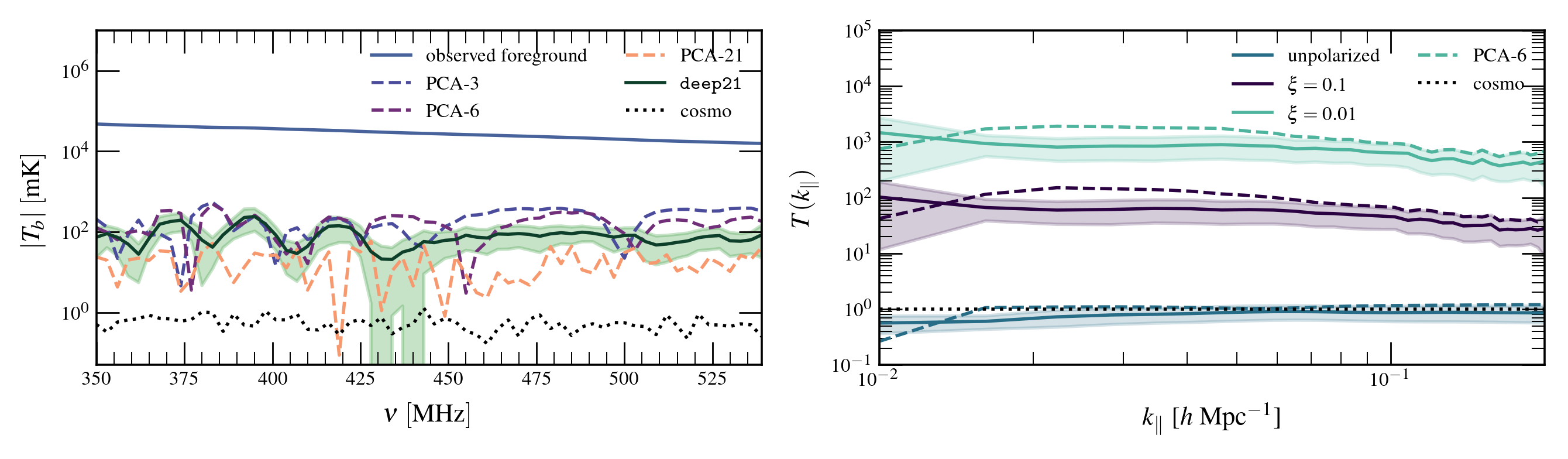}
    \caption{Foreground cleaning errors due to 1\% galactic synchrotron polarization leakage. (\textit{Left}) \deeep{} temperature recovery compared with several PCA subtractions for $\xi_{\rm polar} = 0.01$. Removing as many as 21 PCA components does not resolve the cosmological signal in the presence of polarization leakage. (\textit{Right}) Transfer function for \deeep{} recovery of radial power spectrum recovery for various values of $\xi_{\rm polar}$. Here the PCA-6 subtraction is shown for the given simulation as a dashed line, and shaded contours show $\pm 2\sigma_w$ \deeep{} estimates. Shrinking the polarization correlation length makes leaked galactic synchrotron foregrounds behave similarly to the cosmological signal in frequency, rendering blind methods and \deeep{} ineffective.}
    \label{fig:polar-comp}
\end{figure}

We also tested simulations contaminated by a galactic synchrotron polarization leakage of 1\%. Polarized foregrounds due to the Milky Way's magnetic effects could wreak a catastrophic effect on signal recovered using blind techniques, since poorly-understood polarization leakage could induce foregrounds to interfere with cosmological signal modes \citep{Alonso_sim_2014, LiuShaw20}. To motivate a follow-up study, we enabled galactic synchrotron polarization within the \texttt{CRIME} simulation package, varying the polarized correlation length $\xi_{\rm polar}$, and cleaned the resulting observed maps with our technique. We recovered substantial differences in performance for PCA-6 and \deeep{}, as shown in Figure \ref{fig:polar-comp}. Reducing the correlation length, $\xi_{\rm polar}$ makes leaked galactic synchrotron foreground emission behave similarly to the cosmological signal in frequency (see lefthand plot for $\xi_{\rm polar} = 0.01$, and Figure 8 in \cite{Alonso_sim_2014}). The PCA subtraction fails since synchrotron foregrounds can no longer be smoothly resolved from the choppy cosmological signal. This is shown formally in \cite{Alonso_pca_2014}, Figure 1, where decreasing polarization correlation length spreads foreground contamination across diagonalized PCA eigenvalues, making it more difficult to know when foregrounds have been successfully removed. As expected, \deeep{} does not generalize well to polarized foregrounds, since PCA-3 preprocessing is unable to remove the 1\% synchrotron leakage. We do, however, note that \deeep{} does produce a slight improvement in radial power spectrum recovery (Figure \ref{fig:polar-comp}, right-hand side). 

For a more complete understanding of polarized foregrounds and \deeep{}'s effectiveness in removing them, a follow-up study is warranted. Existing codes such as \texttt{Hammurabi} \citep{hammurabi-gal-fg} make use of detailed three-dimensional Milky Way simulations to model the magnetic fields responsible for polarization leakage. Training \deeep{} on these detailed foregrounds is a necessary next step for real data-preparedness.


\begin{table}[htb]
\centering
\adjustbox{width=\textwidth, center}{
\begin{tabular}{cccccccccc}

\toprule
{} & MSE  & {${\rm T}(\ell)$} & {${\rm T}(k_\parallel)$} & {$\rho_{\rm res}(\ell)$} & {$\rho_{\rm res}(k_\parallel)$} & {${\rm T}(\ell)$} & {${\rm T}(k_\parallel)$} & {$\rho_{\rm res}(\ell)$} & {$\rho_{\rm res}(k_\parallel)$} \\

 {} &       (global)        & $\ell=50$ & $k_\parallel = 0.02$ & $\ell=50$ & $k_\parallel = 0.02$  & $\ell=550$ & $k_\parallel = 0.15$  & $\ell=550$ & $k_\parallel = 0.15$  \\
\midrule
\textbf{test phase}                  &                  &                  &                        &                        &                               &                  &                        &                        &                               \\
$\texttt{deep21}$           &  0.877$\pm$0.156 &  0.899$\pm$0.186 &        0.728$\pm$0.087 &        3.857$\pm$3.178 &               1.099$\pm$0.307 &  0.848$\pm$0.085 &        0.868$\pm$0.088 &         1.099$\pm$0.13 &               0.648$\pm$0.115 \\
PCA-6                       &            17.15 &            0.611 &                  1.095 &                    8.2 &                         1.184 &            0.814 &                  1.195 &                  2.417 &                         1.005 \\
\midrule
$\boldsymbol{\beta = 2.3}$  &                  &                  &                        &                        &                               &                  &                        &                        &                               \\
$\texttt{deep21}$           &  0.872$\pm$0.152 &  0.895$\pm$0.212 &        0.728$\pm$0.088 &        4.545$\pm$4.414 &                 1.076$\pm$0.3 &  0.839$\pm$0.086 &        0.867$\pm$0.088 &        1.063$\pm$0.135 &               0.649$\pm$0.115 \\
PCA-6                       &            17.15 &            0.542 &                  1.093 &                  11.49 &                         1.183 &            0.773 &                  1.194 &                  2.397 &                         1.005 \\
\midrule
$\boldsymbol{\beta = 4.3}$  &                  &                  &                        &                        &                               &                  &                        &                        &                               \\
$\texttt{deep21}$           &  1.044$\pm$0.289 &   0.977$\pm$0.19 &        0.738$\pm$0.081 &         5.83$\pm$5.094 &                1.297$\pm$0.46 &  0.853$\pm$0.114 &        0.874$\pm$0.103 &        1.344$\pm$0.291 &               0.666$\pm$0.132 \\
PCA-6                       &            17.16 &            0.607 &                  1.092 &                  9.025 &                          1.21 &            0.763 &                  1.195 &                  2.805 &                         1.006 \\
\midrule
$\boldsymbol{\alpha = 1.3}$ &                  &                  &                        &                        &                               &                  &                        &                        &                               \\
$\texttt{deep21}$           &  1.764$\pm$1.056 &   1.57$\pm$0.505 &        0.844$\pm$0.081 &        23.761$\pm$23.8 &               1.909$\pm$0.785 &  0.867$\pm$0.111 &         0.87$\pm$0.114 &        1.588$\pm$0.321 &               0.691$\pm$0.143 \\
PCA-6                       &            17.15 &            0.573 &                  1.058 &                  11.35 &                         1.298 &            0.781 &                  1.195 &                  2.707 &                         1.005 \\
\midrule
$\boldsymbol{\alpha = 3.3}$ &                  &                  &                        &                        &                               &                  &                        &                        &                               \\
$\texttt{deep21}$           &  0.871$\pm$0.152 &  0.907$\pm$0.229 &        0.727$\pm$0.088 &        3.135$\pm$3.146 &                 1.073$\pm$0.3 &  0.823$\pm$0.083 &        0.867$\pm$0.088 &        1.164$\pm$0.151 &               0.649$\pm$0.114 \\
PCA-6                       &            17.15 &            0.534 &                  1.095 &                  7.945 &                         1.186 &            0.755 &                  1.194 &                  2.536 &                         1.005 \\
\bottomrule
\end{tabular}}
\caption{Summary of foreground cleaning results. All residuals and MSE metrics are normalized to the corresponding statistic computed for observational noise generated with $\alpha_{\rm noise} = 0.25$. Angular power spectra are computed for a slice at $\nu = 357 \rm MHz$. Uncertainty intervals for \deeep{} were computed for $\pm 2 \sigma_w$ for each statistic. }\label{tab:results}
\end{table}

\begin{figure}[htpb]
    \centering
    \includegraphics[width=\textwidth]{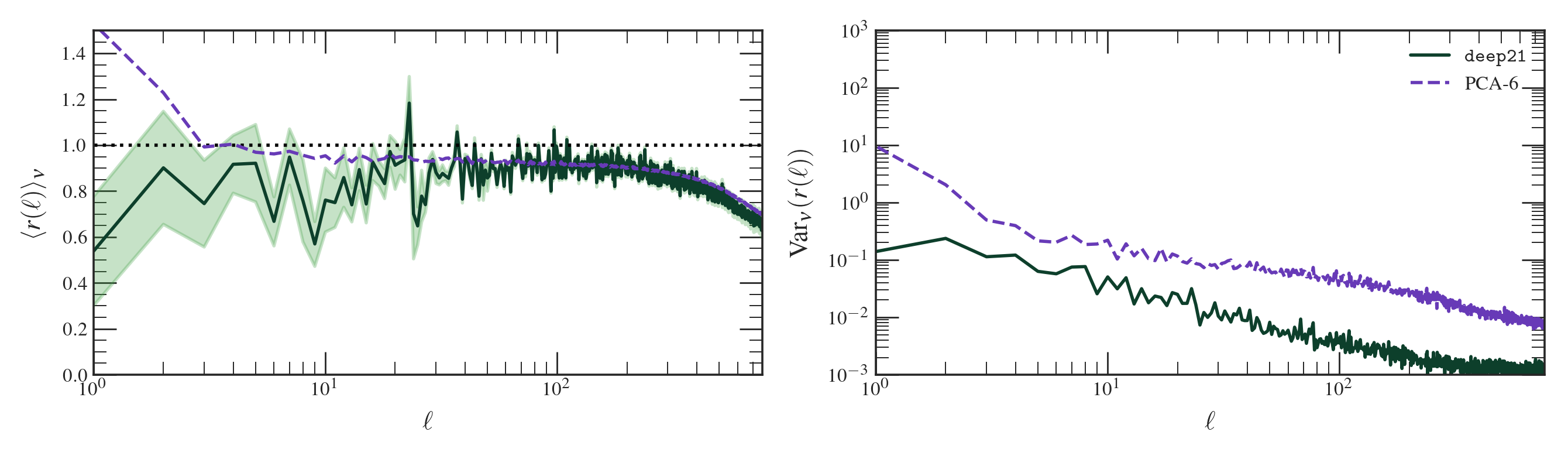}
    \caption{{Mean (\textit{left}) and variance (\textit{right}) of angular power spectrum correlation, $r(\ell)$ computed over frequencies in the first comoving shell for PCA (purple) and \deeep{} methods (green, shaded $\pm 2\sigma_w$). The \deeep{} ensemble's correlation with target signal is consistent with one at all but the smallest scales, showing a distinct improvement at large scales. The variance plot shows that the network method is over an order of magnitude more consistent in capturing map phase information than the more variable PCA as a function of scale.}}
    \label{fig:var-cls}
\end{figure}
\section{Conclusions}\label{sec:conclusions}
In this study we developed a deep learning-based method to improve foreground cleaning techniques for
single-dish 21cm cosmology. Our method outputs clean intensity maps instead of derived summary
statistics. Training an ensemble of independent UNet architectures on simulated foregrounds and
cosmological signal resulted in the improved recovery of intensity maps and angular and radial power
spectra. {Figure \ref{fig:var-cls}) summarises the network's performance in the radial direction. The
lefthand panel shows that on average over frequency, \deeep{}'s $r(\ell)$ is of order 1, offering a
distinct improvement on the largest scales. The righthand plot demonstrates that \deeep{} is over an
order of magnitude more consistent in capturing phase information over frequency than PCA-6.} In
addition, the ensemble method provides an estimate of uncertainty for the summary statistics of
resulting maps. We show that
\deeep{} effectively marginalizes out observational noise at small angular and radial scales,
demonstrating a marked improvement in the sensitivity limit of foreground subtraction over PCA. This
suggests that a learning separation approach could be hardened against instrument effects, presenting a
significant advantage over blind methods. We show that \deeep{} is sensitive to foreground physics by
changing test data simulation
parameters, meaning deep networks trained on more detailed (and varied) simulations will likely be
able to effectively remove foregrounds in absence of a formal foreground likelihood. We also
investigated \deeep{}'s failure modes, namely in the presence of galactic synchrotron polarization
leakage. Improved networks trained on polarized foregrounds (including those which require no PCA
preprocessing) will be the subject of a future work to mitigate these effects on radio map retrieval.
Our method demonstrates that cosmological analyses on previously irretrievable 21cm intensity maps
may be possible in an observational setting.

\subsection{Future Work}
The methods outlined here utilize a simulation-based deep learning method to retrieve intensity maps for radio cosmology. These techniques pave the way for more fundamental studies of the 21cm signal captured by upcoming SKA experiments, as well as future studies with even higher resolution. The ability to retrieve intensity maps will allow future studies to probe structure formation and EoR physics beyond the power spectrum statistic.

However, before \deeep{} can be applied to real 21cm data, several key aspects should be addressed in  follow-up studies:
\begin{itemize}
    \item {Rigorous quantification of network errors: although \deeep{} achieves a more consistent foreground cleaning than blind methods, the error in the network method is still not fully understood. A way to quantify this effect is to incorporate a \deeep{}-like cleaning method in a simulation-based inference pipeline for compressed cosmological parameters \citep[see e.g.][]{Alsing_2019, Charnock_2018}. If foreground cleaning is performed as a step in a simulation pipeline, cleaning method errors can be propagated to parameter measurement using Approximate Bayesian Computational methods.}
    \item {BAO application: a future study might train \deeep{} on cosmological data without BAO oscillations, and then ask the network to clean data with BAO signal. Training a network with a fiducial BAO radius would introduce a bias, but cleaning a map with a BAO-free network might be trained to leave an unbiased BAO residual signal in the predicted map, since BAO signal is expected to be statistically distinct from foreground and cosmological signal \citep{Wyithe_2007, paco-2016}.}
    \item A more realistic noise model: the white noise model considered here, while frequency-dependent, will not extend to more complicated intensity map datasets, such as those obtained via interferometry \citep{Shaw_2014}. Thus the impact of nontrivial noise correlations on network performance must be considered before real signals can be reliably separated. A \deeep{}-like study on systematics would additionally benefit from varying observed sky fraction, as is done in \citet{paco-2016} for 21cm BAO recovery.
    \item Varying astrophysical parameters: here we trained \deeep{} on thousands of input voxels derived from the same fiducial cosmological and foreground parameters. While the network performed relatively well in some generalization cases, \deeep{} would ideally be trained on a range of simulation parameters, like is done by \citet{Pablo_2020}, before asked to clean real data.
    \item Training on polarized foregrounds: We additionally probed a failure mode of the network in the presence of 1\% galactic synchrotron radiation polarization leakage. Here the PCA preprocessing fails to separate the leaked and cosmological signals, making it harder for \deeep{} to pick out the cosmological signal. A follow-up study might additionally train on more realistic polarized simulations, such as those produced by \texttt{Hammurabi}, as well as bypass the need for a blind preprocessing step.
    \item Increasing input sizes: \deeep{}'s input voxel sizes were limited by available GPU memory. With improved deep learning computational resources (or a detailed tiling strategy such as the one employed by \citet{shirleyd3m} for N-Body analyses), Increasing input size will likely improve foreground removal on large scales, since the network will have access to a larger context of information. Ideally, entire maps would only be split into a handful of UNet input units.  An increase in input volume might also allow for a larger frequency range to be assessed, which would aid the network in distinguishing cosmological and polarized signals, as well as improve radial power spectrum recovery.
\end{itemize}


\subsection*{Code Availability}
The code used for training and generation of results is publicly available at \url{https://github.com/tlmakinen/deep21} \faGithub. 
A browser-based tutorial for the experiment and UNet module is available via the accompanying \href{https://colab.research.google.com/drive/1wQnmelM33Qjq-nHeVD9JkTHXER1PAJM0?hl=en#scrollTo=yyqq36iyWJ6g}{Colab notebook} \faGoogle. 

\section{Acknowledgements}
The authors would like to thank the referee for their constructive comments guiding this revision. Many thanks to Nick Carriero and the Flatiron Institute's HPC support team, without whom this work would not be possible. Thank you also to David Alonso for simulation guidance and to Ben Wandelt for helpful discussions. FVN acknowledges funding from the WFIRST program through NNG26PJ30C and NNN12AA01c. The work of SH and DNS has been supported by the Simons Foundation. TLM completed a large portion of this work to satisfy requirements for the Degree of Bachelor of Arts at Princeton University.




\bibliographystyle{plainnat}
\bibliography{bibliography.bib}

\begin{thebibliography}{89}
\providecommand{\natexlab}[1]{#1}
\providecommand{\url}[1]{\texttt{#1}}
\expandafter\ifx\csname urlstyle\endcsname\relax
  \providecommand{\doi}[1]{doi: #1}\else
  \providecommand{\doi}{doi: \begingroup \urlstyle{rm}\Url}\fi

\bibitem[Abdalla and Rawlings(2005)]{Abdalla_2005}
F.~B. Abdalla and S.~Rawlings.
\newblock Probing dark energy with baryonic oscillations and future radio
  surveys of neutral hydrogen.
\newblock \emph{Monthly Notices of the Royal Astronomical Society},
  360\penalty0 (1):\penalty0 27–40, Jun 2005.
\newblock ISSN 1365-2966.
\newblock \doi{10.1111/j.1365-2966.2005.08650.x}.
\newblock URL \url{http://dx.doi.org/10.1111/j.1365-2966.2005.08650.x}.

\bibitem[Alonso et~al.(2014{\natexlab{a}})Alonso, Bull, Ferreira, and
  Santos]{Alonso_pca_2014}
David Alonso, Philip Bull, Pedro~G. Ferreira, and Mário~G. Santos.
\newblock Blind foreground subtraction for intensity mapping experiments.
\newblock \emph{Monthly Notices of the Royal Astronomical Society},
  447\penalty0 (1):\penalty0 400–416, Dec 2014{\natexlab{a}}.
\newblock ISSN 0035-8711.
\newblock \doi{10.1093/mnras/stu2474}.
\newblock URL \url{http://dx.doi.org/10.1093/mnras/stu2474}.

\bibitem[Alonso et~al.(2014{\natexlab{b}})Alonso, Ferreira, and
  Santos]{Alonso_sim_2014}
David Alonso, Pedro~G. Ferreira, and Mario~G. Santos.
\newblock Fast simulations for intensity mapping experiments.
\newblock \emph{Monthly Notices of the Royal Astronomical Society},
  444\penalty0 (4):\penalty0 3183–3197, Sep 2014{\natexlab{b}}.
\newblock ISSN 0035-8711.
\newblock \doi{10.1093/mnras/stu1666}.
\newblock URL \url{http://dx.doi.org/10.1093/mnras/stu1666}.

\bibitem[Alsing et~al.(2019)Alsing, Charnock, Feeney, and Wandelt]{Alsing_2019}
Justin Alsing, Tom Charnock, Stephen Feeney, and Benjamin Wandelt.
\newblock Fast likelihood-free cosmology with neural density estimators and
  active learning.
\newblock \emph{Monthly Notices of the Royal Astronomical Society}, Jul 2019.
\newblock ISSN 1365-2966.
\newblock \doi{10.1093/mnras/stz1960}.
\newblock URL \url{http://dx.doi.org/10.1093/mnras/stz1960}.

\bibitem[Bacon et~al.(2020)Bacon, Battye, Bull, Camera, Ferreira, Harrison,
  Parkinson, Pourtsidou, Santos, and et~al.]{ska2020}
David~J. Bacon, Richard~A. Battye, Philip Bull, Stefano Camera, Pedro~G.
  Ferreira, Ian Harrison, David Parkinson, Alkistis Pourtsidou, Mário~G.
  Santos, and et~al.
\newblock Cosmology with phase 1 of the square kilometre array red book 2018:
  Technical specifications and performance forecasts.
\newblock \emph{Publications of the Astronomical Society of Australia}, 37,
  2020.
\newblock ISSN 1448-6083.
\newblock \doi{10.1017/pasa.2019.51}.
\newblock URL \url{http://dx.doi.org/10.1017/pasa.2019.51}.

\bibitem[Bergstra et~al.(2015)Bergstra, Komer, Eliasmith, Yamins, and
  Cox]{hyperopt}
James Bergstra, Brent Komer, Chris Eliasmith, Dan Yamins, and David Cox.
\newblock Hyperopt: A python library for model selection and hyperparameter
  optimization.
\newblock \emph{Computational Science \& Discovery}, 8:\penalty0 014008, 07
  2015.
\newblock \doi{10.1088/1749-4699/8/1/014008}.

\bibitem[Bernardi et~al.(2009)Bernardi, de~Bruyn, Brentjens, Ciardi, Harker,
  Jelić, Koopmans, Labropoulos, Offringa, Pandey, and et~al.]{Bernardi_2009}
G.~Bernardi, A.~G. de~Bruyn, M.~A. Brentjens, B.~Ciardi, G.~Harker, V.~Jelić,
  L.~V.~E. Koopmans, P.~Labropoulos, A.~Offringa, V.~N. Pandey, and et~al.
\newblock Foregrounds for observations of the cosmological 21 cm line.
\newblock \emph{Astronomy \& Astrophysics}, 500\penalty0 (3):\penalty0
  965–979, Jun 2009.
\newblock ISSN 1432-0746.
\newblock \doi{10.1051/0004-6361/200911627}.
\newblock URL \url{http://dx.doi.org/10.1051/0004-6361/200911627}.

\bibitem[Bernardi et~al.(2010)Bernardi, de~Bruyn, Harker, Brentjens, Ciardi,
  Jelić, Koopmans, Labropoulos, Offringa, Pandey, and et~al.]{Bernardi_2010}
G.~Bernardi, A.~G. de~Bruyn, G.~Harker, M.~A. Brentjens, B.~Ciardi, V.~Jelić,
  L.~V.~E. Koopmans, P.~Labropoulos, A.~Offringa, V.~N. Pandey, and et~al.
\newblock Foregrounds for observations of the cosmological 21 cm line.
\newblock \emph{Astronomy \& Astrophysics}, 522:\penalty0 A67, Nov 2010.
\newblock ISSN 1432-0746.
\newblock \doi{10.1051/0004-6361/200913420}.
\newblock URL \url{http://dx.doi.org/10.1051/0004-6361/200913420}.

\bibitem[{Bull} et~al.(2015){Bull}, {Ferreira}, {Patel}, and
  {Santos}]{bull-late-stage-cosmo-21}
Philip {Bull}, Pedro~G. {Ferreira}, Prina {Patel}, and M{\'a}rio~G. {Santos}.
\newblock {Late-time Cosmology with 21 cm Intensity Mapping Experiments}.
\newblock \emph{\apj}, 803\penalty0 (1):\penalty0 21, April 2015.
\newblock \doi{10.1088/0004-637X/803/1/21}.

\bibitem[{Camera} et~al.(2013){Camera}, {Santos}, {Ferreira}, and
  {Ferramacho}]{camera-2013}
Stefano {Camera}, M{\'a}rio~G. {Santos}, Pedro~G. {Ferreira}, and Lu{\'\i}s
  {Ferramacho}.
\newblock {Cosmology on Ultralarge Scales with Intensity Mapping of the Neutral
  Hydrogen 21 cm Emission: Limits on Primordial Non-Gaussianity}.
\newblock \emph{\prl}, 111\penalty0 (17):\penalty0 171302, October 2013.
\newblock \doi{10.1103/PhysRevLett.111.171302}.

\bibitem[{Chang} et~al.(2010){Chang}, {Pen}, {Bandura}, and
  {Peterson}]{chang-pca-real}
Tzu-Ching {Chang}, Ue-Li {Pen}, Kevin {Bandura}, and Jeffrey~B. {Peterson}.
\newblock {Hydrogen 21-cm Intensity Mapping at redshift 0.8}.
\newblock \emph{arXiv e-prints}, art. arXiv:1007.3709, July 2010.

\bibitem[Chardin et~al.(2019)Chardin, Uhlrich, Aubert, Deparis, Gillet, Ocvirk,
  and Lewis]{Chardin_2019}
Jonathan Chardin, Grégoire Uhlrich, Dominique Aubert, Nicolas Deparis, Nicolas
  Gillet, Pierre Ocvirk, and Joseph Lewis.
\newblock A deep learning model to emulate simulations of cosmic reionization.
\newblock \emph{Monthly Notices of the Royal Astronomical Society},
  490\penalty0 (1):\penalty0 1055–1065, Sep 2019.
\newblock ISSN 1365-2966.
\newblock \doi{10.1093/mnras/stz2605}.
\newblock URL \url{http://dx.doi.org/10.1093/mnras/stz2605}.

\bibitem[Charnock et~al.(2018)Charnock, Lavaux, and Wandelt]{Charnock_2018}
Tom Charnock, Guilhem Lavaux, and Benjamin~D. Wandelt.
\newblock Automatic physical inference with information maximizing neural
  networks.
\newblock \emph{Physical Review D}, 97\penalty0 (8), Apr 2018.
\newblock ISSN 2470-0029.
\newblock \doi{10.1103/physrevd.97.083004}.
\newblock URL \url{http://dx.doi.org/10.1103/PhysRevD.97.083004}.

\bibitem[Charnock et~al.(2020)Charnock, Perreault-Levasseur, and
  Lanusse]{charnock2020bayesian}
Tom Charnock, Laurence Perreault-Levasseur, and François Lanusse.
\newblock Bayesian neural networks, 2020.

\bibitem[{Cheng} et~al.(2018){Cheng}, {Parsons}, {Kolopanis}, {Jacobs}, {Liu},
  {Kohn}, {Aguirre}, {Pober}, {Ali}, {Bernardi}, {Bradley}, {Carilli},
  {DeBoer}, {Dexter}, {Dillon}, {Klima}, {MacMahon}, {Moore}, {Nunhokee},
  {Walbrugh}, and {Walker}]{2018ApJ_signalloss_inter}
Carina {Cheng}, Aaron~R. {Parsons}, Matthew {Kolopanis}, Daniel~C. {Jacobs},
  Adrian {Liu}, Saul~A. {Kohn}, James~E. {Aguirre}, Jonathan~C. {Pober},
  Zaki~S. {Ali}, Gianni {Bernardi}, Richard~F. {Bradley}, Chris~L. {Carilli},
  David~R. {DeBoer}, Matthew~R. {Dexter}, Joshua~S. {Dillon}, Pat {Klima},
  David H.~E. {MacMahon}, David~F. {Moore}, Chuneeta~D. {Nunhokee}, William~P.
  {Walbrugh}, and Andre {Walker}.
\newblock {Characterizing Signal Loss in the 21 cm Reionization Power Spectrum:
  A Revised Study of PAPER-64}.
\newblock \emph{\apj}, 868\penalty0 (1):\penalty0 26, November 2018.
\newblock \doi{10.3847/1538-4357/aae833}.

\bibitem[{Cohen} et~al.(2004){Cohen}, {R{\"o}ttgering}, {Jarvis}, {Kassim}, and
  {Lazio}]{cohen_2004}
A.~S. {Cohen}, H.~J.~A. {R{\"o}ttgering}, M.~J. {Jarvis}, N.~E. {Kassim}, and
  T.~J.~W. {Lazio}.
\newblock {A Deep, High-Resolution Survey at 74 MHz}.
\newblock \emph{\apjs}, 150\penalty0 (2):\penalty0 417--430, February 2004.
\newblock \doi{10.1086/380783}.

\bibitem[Cunnington et~al.(2019)Cunnington, Wolz, Pourtsidou, and
  Bacon]{Cunnington_2019}
Steven Cunnington, Laura Wolz, Alkistis Pourtsidou, and David Bacon.
\newblock Impact of foregrounds on h i intensity mapping cross-correlations
  with optical surveys.
\newblock \emph{Monthly Notices of the Royal Astronomical Society},
  488\penalty0 (4):\penalty0 5452–5472, Jul 2019.
\newblock ISSN 1365-2966.
\newblock \doi{10.1093/mnras/stz1916}.
\newblock URL \url{http://dx.doi.org/10.1093/mnras/stz1916}.

\bibitem[de~Bruyn et~al.(2006)de~Bruyn, Katgert, Haverkorn, and
  Schnitzeler]{deBruyn2006}
A.~de~Bruyn, P.~Katgert, Marijke Haverkorn, and D.~Schnitzeler.
\newblock Radio polarization and rm structure at high galactic latitudes.
\newblock \emph{Astronomische Nachrichten}, 327:\penalty0 487 -- 490, 06 2006.
\newblock \doi{10.1002/asna.200610566}.

\bibitem[{de Oliveira-Costa} et~al.(2008){de Oliveira-Costa}, {Tegmark},
  {Gaensler}, {Jonas}, {Landecker}, and {Reich}]{pca_oliveira2008}
Ang{\'e}lica {de Oliveira-Costa}, Max {Tegmark}, B.~M. {Gaensler}, Justin
  {Jonas}, T.~L. {Landecker}, and Patricia {Reich}.
\newblock {A model of diffuse Galactic radio emission from 10 MHz to 100 GHz}.
\newblock \emph{\mnras}, 388\penalty0 (1):\penalty0 247--260, July 2008.
\newblock \doi{10.1111/j.1365-2966.2008.13376.x}.

\bibitem[{Field}(1958)]{field_58}
George~B. {Field}.
\newblock {Excitation of the Hydrogen 21-CM Line}.
\newblock \emph{Proceedings of the IRE}, 46:\penalty0 240--250, January 1958.
\newblock \doi{10.1109/JRPROC.1958.286741}.

\bibitem[{Field}(1959)]{field_59}
George~B. {Field}.
\newblock {The Spin Temperature of Intergalactic Neutral Hydrogen.}
\newblock \emph{\apj}, 129:\penalty0 536, May 1959.
\newblock \doi{10.1086/146653}.

\bibitem[{Fort} et~al.(2019){Fort}, {Hu}, and
  {Lakshminarayanan}]{deep_ens_loss_land}
Stanislav {Fort}, Huiyi {Hu}, and Balaji {Lakshminarayanan}.
\newblock {Deep Ensembles: A Loss Landscape Perspective}.
\newblock \emph{arXiv e-prints}, art. arXiv:1912.02757, December 2019.

\bibitem[{Furlanetto} et~al.(2006){Furlanetto}, {Oh}, and
  {Briggs}]{Furlanetto06}
Steven~R. {Furlanetto}, S.~Peng {Oh}, and Frank~H. {Briggs}.
\newblock {Cosmology at low frequencies: The 21 cm transition and the
  high-redshift Universe}.
\newblock \emph{\physrep}, 433\penalty0 (4-6):\penalty0 181--301, October 2006.
\newblock \doi{10.1016/j.physrep.2006.08.002}.

\bibitem[Gillet et~al.(2019)Gillet, Mesinger, Greig, Liu, and
  Ucci]{deeplearn-eor-Gillet_2019}
Nicolas Gillet, Andrei Mesinger, Bradley Greig, Adrian Liu, and Graziano Ucci.
\newblock Deep learning from 21-cm tomography of the cosmic dawn and
  reionization.
\newblock \emph{Monthly Notices of the Royal Astronomical Society}, Jan 2019.
\newblock ISSN 1365-2966.
\newblock \doi{10.1093/mnras/stz010}.
\newblock URL \url{http://dx.doi.org/10.1093/mnras/stz010}.

\bibitem[Gleser et~al.(2008)Gleser, Nusser, and Benson]{Gleser_2008}
Liron Gleser, Adi Nusser, and Andrew~J. Benson.
\newblock Decontamination of cosmological 21-cm maps.
\newblock \emph{Monthly Notices of the Royal Astronomical Society},
  391\penalty0 (1):\penalty0 383–398, Nov 2008.
\newblock ISSN 1365-2966.
\newblock \doi{10.1111/j.1365-2966.2008.13897.x}.
\newblock URL \url{http://dx.doi.org/10.1111/j.1365-2966.2008.13897.x}.

\bibitem[Glorot and Bengio(2010)]{Glorot10understandingthe}
Xavier Glorot and Yoshua Bengio.
\newblock Understanding the difficulty of training deep feedforward neural
  networks.
\newblock In \emph{In Proceedings of the International Conference on Artificial
  Intelligence and Statistics (AISTATS’10). Society for Artificial
  Intelligence and Statistics}, 2010.

\bibitem[Goodfellow et~al.(2016)Goodfellow, Bengio, and
  Courville]{GoodBengCour16}
Ian~J. Goodfellow, Yoshua Bengio, and Aaron Courville.
\newblock \emph{Deep Learning}.
\newblock MIT Press, Cambridge, MA, USA, 2016.
\newblock \url{http://www.deeplearningbook.org}.

\bibitem[{G{\'o}rski} et~al.(2005){G{\'o}rski}, {Hivon}, {Banday}, {Wandelt},
  {Hansen}, {Reinecke}, and {Bartelmann}]{healpix}
K.~M. {G{\'o}rski}, E.~{Hivon}, A.~J. {Banday}, B.~D. {Wandelt}, F.~K.
  {Hansen}, M.~{Reinecke}, and M.~{Bartelmann}.
\newblock {HEALPix: A Framework for High-Resolution Discretization and Fast
  Analysis of Data Distributed on the Sphere}.
\newblock \emph{\apj}, 622:\penalty0 759--771, April 2005.
\newblock \doi{10.1086/427976}.

\bibitem[{Hall} et~al.(2013){Hall}, {Bonvin}, and {Challinor}]{hall-2013}
Alex {Hall}, Camille {Bonvin}, and Anthony {Challinor}.
\newblock {Testing general relativity with 21-cm intensity mapping}.
\newblock \emph{\prd}, 87\penalty0 (6):\penalty0 064026, March 2013.
\newblock \doi{10.1103/PhysRevD.87.064026}.

\bibitem[{Haslam} et~al.(1982){Haslam}, {Salter}, {Stoffel}, and
  {Wilson}]{haslam1982}
C.~G.~T. {Haslam}, C.~J. {Salter}, H.~{Stoffel}, and W.~E. {Wilson}.
\newblock {A 408 MHz all-sky continuum survey. II. The atlas of contour maps.}
\newblock \emph{\aaps}, 47:\penalty0 1--143, January 1982.

\bibitem[He et~al.(2019)He, Li, Feng, Ho, Ravanbakhsh, Chen, and
  P{\'o}czos]{shirleyd3m}
Siyu He, Yin Li, Yu~Feng, Shirley Ho, Siamak Ravanbakhsh, Wei Chen, and
  Barnab{\'a}s P{\'o}czos.
\newblock Learning to predict the cosmological structure formation.
\newblock \emph{Proceedings of the National Academy of Sciences}, 116\penalty0
  (28):\penalty0 13825--13832, 2019.
\newblock ISSN 0027-8424.
\newblock \doi{10.1073/pnas.1821458116}.
\newblock URL \url{https://www.pnas.org/content/116/28/13825}.

\bibitem[Jelić et~al.(2008)Jelić, Zaroubi, Labropoulos, Thomas, Bernardi,
  Brentjens, de~Bruyn, Ciardi, Harker, Koopmans, and et~al.]{Jeli__2008}
V.~Jelić, S.~Zaroubi, P.~Labropoulos, R.~M. Thomas, G.~Bernardi, M.~A.
  Brentjens, A.~G. de~Bruyn, B.~Ciardi, G.~Harker, L.~V.~E. Koopmans, and
  et~al.
\newblock Foreground simulations for the lofar-epoch of reionization
  experiment.
\newblock \emph{Monthly Notices of the Royal Astronomical Society},
  389\penalty0 (3):\penalty0 1319–1335, Sep 2008.
\newblock ISSN 1365-2966.
\newblock \doi{10.1111/j.1365-2966.2008.13634.x}.
\newblock URL \url{http://dx.doi.org/10.1111/j.1365-2966.2008.13634.x}.

\bibitem[{Kohl} et~al.(2018){Kohl}, {Romera-Paredes}, {Meyer}, {De Fauw},
  {Ledsam}, {Maier-Hein}, {Eslami}, {Jimenez Rezende}, and
  {Ronneberger}]{prob-unet}
Simon A.~A. {Kohl}, Bernardino {Romera-Paredes}, Clemens {Meyer}, Jeffrey {De
  Fauw}, Joseph~R. {Ledsam}, Klaus~H. {Maier-Hein}, S.~M.~Ali {Eslami}, Danilo
  {Jimenez Rezende}, and Olaf {Ronneberger}.
\newblock {A Probabilistic U-Net for Segmentation of Ambiguous Images}.
\newblock \emph{arXiv e-prints}, art. arXiv:1806.05034, June 2018.

\bibitem[Kwon et~al.(2020)Kwon, Hong, and Park]{Kwon-deeplearn-21_2020}
Yungi Kwon, Sungwook~E. Hong, and Inkyu Park.
\newblock Deep-learning study of the 21-cm differential brightness temperature
  during the epoch of reionization.
\newblock \emph{Journal of the Korean Physical Society}, 77\penalty0
  (1):\penalty0 49–59, Jul 2020.
\newblock ISSN 1976-8524.
\newblock \doi{10.3938/jkps.77.49}.
\newblock URL \url{http://dx.doi.org/10.3938/jkps.77.49}.

\bibitem[Lakshminarayanan et~al.(2017)Lakshminarayanan, Pritzel, and
  Blundell]{deep_ens}
Balaji Lakshminarayanan, Alexander Pritzel, and Charles Blundell.
\newblock Simple and scalable predictive uncertainty estimation using deep
  ensembles.
\newblock In I.~Guyon, U.~V. Luxburg, S.~Bengio, H.~Wallach, R.~Fergus,
  S.~Vishwanathan, and R.~Garnett, editors, \emph{Advances in Neural
  Information Processing Systems 30}, pages 6402--6413. Curran Associates,
  Inc., 2017.

\bibitem[Li et~al.(2019)Li, Xu, Ma, Zhu, Hu, Zhu, Gu, Shan, Zhu, and
  Wu]{Li_2019}
Weitian Li, Haiguang Xu, Zhixian Ma, Ruimin Zhu, Dan Hu, Zhenghao Zhu, Junhua
  Gu, Chenxi Shan, Jie Zhu, and Xiang-Ping Wu.
\newblock Separating the eor signal with a convolutional denoising autoencoder:
  a deep-learning-based method.
\newblock \emph{Monthly Notices of the Royal Astronomical Society},
  485\penalty0 (2):\penalty0 2628–2637, Feb 2019.
\newblock ISSN 1365-2966.
\newblock \doi{10.1093/mnras/stz582}.
\newblock URL \url{http://dx.doi.org/10.1093/mnras/stz582}.

\bibitem[List and Lewis(2020)]{List_2020}
Florian List and Geraint~F Lewis.
\newblock A unified framework for 21 cm tomography sample generation and
  parameter inference with progressively growing gans.
\newblock \emph{Monthly Notices of the Royal Astronomical Society},
  493\penalty0 (4):\penalty0 5913–5927, Feb 2020.
\newblock ISSN 1365-2966.
\newblock \doi{10.1093/mnras/staa523}.
\newblock URL \url{http://dx.doi.org/10.1093/mnras/staa523}.

\bibitem[{Liu} and {Shaw}(2020)]{LiuShaw20}
Adrian {Liu} and J.~Richard {Shaw}.
\newblock {Data Analysis for Precision 21 cm Cosmology}.
\newblock \emph{\pasp}, 132\penalty0 (1012):\penalty0 062001, June 2020.
\newblock \doi{10.1088/1538-3873/ab5bfd}.

\bibitem[Liu and Tegmark(2011)]{Liu_2011}
Adrian Liu and Max Tegmark.
\newblock A method for 21 cm power spectrum estimation in the presence of
  foregrounds.
\newblock \emph{Physical Review D}, 83\penalty0 (10), May 2011.
\newblock ISSN 1550-2368.
\newblock \doi{10.1103/physrevd.83.103006}.
\newblock URL \url{http://dx.doi.org/10.1103/PhysRevD.83.103006}.

\bibitem[Liu and Tegmark(2012)]{tegmark-liu12}
Adrian Liu and Max Tegmark.
\newblock {How well can we measure and understand foregrounds with 21-cm
  experiments?}
\newblock \emph{Monthly Notices of the Royal Astronomical Society},
  419\penalty0 (4):\penalty0 3491--3504, 01 2012.
\newblock ISSN 0035-8711.
\newblock \doi{10.1111/j.1365-2966.2011.19989.x}.
\newblock URL \url{https://doi.org/10.1111/j.1365-2966.2011.19989.x}.

\bibitem[Liu et~al.(2009)Liu, Tegmark, Bowman, Hewitt, and
  Zaldarriaga]{Liu_2009}
Adrian Liu, Max Tegmark, Judd Bowman, Jacqueline Hewitt, and Matias
  Zaldarriaga.
\newblock An improved method for 21-cm foreground removal.
\newblock \emph{Monthly Notices of the Royal Astronomical Society},
  398\penalty0 (1):\penalty0 401–406, Sep 2009.
\newblock ISSN 1365-2966.
\newblock \doi{10.1111/j.1365-2966.2009.15156.x}.
\newblock URL \url{http://dx.doi.org/10.1111/j.1365-2966.2009.15156.x}.

\bibitem[Loshchilov and Hutter(2019)]{adamw}
Ilya Loshchilov and Frank Hutter.
\newblock Decoupled weight decay regularization, 2019.

\bibitem[{Madau} et~al.(1997){Madau}, {Meiksin}, and {Rees}]{Madau97}
Piero {Madau}, Avery {Meiksin}, and Martin~J. {Rees}.
\newblock {21 Centimeter Tomography of the Intergalactic Medium at High
  Redshift}.
\newblock \emph{\apj}, 475\penalty0 (2):\penalty0 429--444, February 1997.
\newblock \doi{10.1086/303549}.

\bibitem[{Mangena} et~al.(2020){Mangena}, {Hassan}, and {Santos}]{Hassan2020}
Tumelo {Mangena}, Sultan {Hassan}, and Mario~G. {Santos}.
\newblock {Constraining the reionization history using deep learning from 21cm
  tomography with the Square Kilometre Array}.
\newblock \emph{\mnras}, March 2020.
\newblock \doi{10.1093/mnras/staa750}.

\bibitem[Mangena et~al.(2020{\natexlab{a}})Mangena, Hassan, and
  Santos]{Mangena_2020}
Tumelo Mangena, Sultan Hassan, and Mario~G Santos.
\newblock Constraining the reionization history using deep learning from 21-cm
  tomography with the square kilometre array.
\newblock \emph{Monthly Notices of the Royal Astronomical Society},
  494\penalty0 (1):\penalty0 600–606, Mar 2020{\natexlab{a}}.
\newblock ISSN 1365-2966.
\newblock \doi{10.1093/mnras/staa750}.
\newblock URL \url{http://dx.doi.org/10.1093/mnras/staa750}.

\bibitem[Mangena et~al.(2020{\natexlab{b}})Mangena, Hassan, and
  Santos]{deeplearn-21-Mangena_2020}
Tumelo Mangena, Sultan Hassan, and Mario~G Santos.
\newblock Constraining the reionization history using deep learning from 21-cm
  tomography with the square kilometre array.
\newblock \emph{Monthly Notices of the Royal Astronomical Society},
  494\penalty0 (1):\penalty0 600–606, Mar 2020{\natexlab{b}}.
\newblock ISSN 1365-2966.
\newblock \doi{10.1093/mnras/staa750}.
\newblock URL \url{http://dx.doi.org/10.1093/mnras/staa750}.

\bibitem[Masui et~al.(2013)Masui, Switzer, Banavar, Bandura, Blake, Calin,
  Chang, Chen, Li, Liao, and et~al.]{Masui_2013}
K.~W. Masui, E.~R. Switzer, N.~Banavar, K.~Bandura, C.~Blake, L.-M. Calin,
  T.-C. Chang, X.~Chen, Y.-C. Li, Y.-W. Liao, and et~al.
\newblock Measurement of 21 cm brightness fluctuations atvz ∼ 0.8 in
  cross-correlation.
\newblock \emph{The Astrophysical Journal}, 763\penalty0 (1):\penalty0 L20, Jan
  2013.
\newblock ISSN 2041-8213.
\newblock \doi{10.1088/2041-8205/763/1/l20}.
\newblock URL \url{http://dx.doi.org/10.1088/2041-8205/763/1/L20}.

\bibitem[Matteo et~al.(2002)Matteo, Perna, Abel, and Rees]{Di_Matteo_2002}
Tiziana~Di Matteo, Rosalba Perna, Tom Abel, and Martin~J. Rees.
\newblock Radio foregrounds for the 21 centimeter tomography of the neutral
  intergalactic medium at high redshifts.
\newblock \emph{The Astrophysical Journal}, 564\penalty0 (2):\penalty0
  576--580, jan 2002.
\newblock \doi{10.1086/324293}.
\newblock URL \url{https://doi.org/10.1086%2F324293}.

\bibitem[Mhaskar and Poggio(2016)]{network-depth-mhaskar2016deep}
Hrushikesh Mhaskar and Tomaso Poggio.
\newblock Deep vs. shallow networks : An approximation theory perspective,
  2016.

\bibitem[{Monsalve} et~al.(2017){Monsalve}, {Rogers}, {Bowman}, and
  {Mozdzen}]{edges2017}
Raul~A. {Monsalve}, Alan E.~E. {Rogers}, Judd~D. {Bowman}, and Thomas~J.
  {Mozdzen}.
\newblock {Results from EDGES High-band. I. Constraints on Phenomenological
  Models for the Global 21 cm Signal}.
\newblock \emph{\apj}, 847\penalty0 (1):\penalty0 64, September 2017.
\newblock \doi{10.3847/1538-4357/aa88d1}.

\bibitem[{Monsalve} et~al.(2018){Monsalve}, {Greig}, {Bowman}, {Mesinger},
  {Rogers}, {Mozdzen}, {Kern}, and {Mahesh}]{edges-2018}
Raul~A. {Monsalve}, Bradley {Greig}, Judd~D. {Bowman}, Andrei {Mesinger}, Alan
  E.~E. {Rogers}, Thomas~J. {Mozdzen}, Nicholas~S. {Kern}, and Nivedita
  {Mahesh}.
\newblock {Results from EDGES High-band. II. Constraints on Parameters of Early
  Galaxies}.
\newblock \emph{\apj}, 863\penalty0 (1):\penalty0 11, August 2018.
\newblock \doi{10.3847/1538-4357/aace54}.

\bibitem[Moore et~al.(2013)Moore, Aguirre, Parsons, Jacobs, and
  Pober]{Moore_2013}
David~F. Moore, James~E. Aguirre, Aaron~R. Parsons, Daniel~C. Jacobs, and
  Jonathan~C. Pober.
\newblock The effects of polarized foregrounds on 21 cm epoch of reionization
  power spectrum measurements.
\newblock \emph{The Astrophysical Journal}, 769\penalty0 (2):\penalty0 154, May
  2013.
\newblock ISSN 1538-4357.
\newblock \doi{10.1088/0004-637x/769/2/154}.
\newblock URL \url{http://dx.doi.org/10.1088/0004-637X/769/2/154}.

\bibitem[{Morales} and {Wyithe}(2010)]{MoralesWyithe10}
Miguel~F. {Morales} and J.~Stuart~B. {Wyithe}.
\newblock {Reionization and Cosmology with 21-cm Fluctuations}.
\newblock \emph{\araa}, 48:\penalty0 127--171, September 2010.
\newblock \doi{10.1146/annurev-astro-081309-130936}.

\bibitem[Morales et~al.(2006)Morales, Bowman, and Hewitt]{Morales_2006}
Miguel~F. Morales, Judd~D. Bowman, and Jacqueline~N. Hewitt.
\newblock Improving foreground subtraction in statistical observations of 21 cm
  emission from the epoch of reionization.
\newblock \emph{The Astrophysical Journal}, 648\penalty0 (2):\penalty0
  767–773, Sep 2006.
\newblock ISSN 1538-4357.
\newblock \doi{10.1086/506135}.
\newblock URL \url{http://dx.doi.org/10.1086/506135}.

\bibitem[Pedregosa et~al.(2011)Pedregosa, Varoquaux, Gramfort, Michel, Thirion,
  Grisel, Blondel, Prettenhofer, Weiss, Dubourg, Vanderplas, Passos,
  Cournapeau, Brucher, Perrot, and Duchesnay]{scikit-learn}
F.~Pedregosa, G.~Varoquaux, A.~Gramfort, V.~Michel, B.~Thirion, O.~Grisel,
  M.~Blondel, P.~Prettenhofer, R.~Weiss, V.~Dubourg, J.~Vanderplas, A.~Passos,
  D.~Cournapeau, M.~Brucher, M.~Perrot, and E.~Duchesnay.
\newblock Scikit-learn: Machine learning in {P}ython.
\newblock \emph{Journal of Machine Learning Research}, 12:\penalty0 2825--2830,
  2011.

\bibitem[Peng~Oh and Mack(2003)]{Peng_Oh_2003}
S.~Peng~Oh and Katherine~J. Mack.
\newblock Foregrounds for 21-cm observations of neutral gas at high redshift.
\newblock \emph{Monthly Notices of the Royal Astronomical Society},
  346\penalty0 (3):\penalty0 871–877, Dec 2003.
\newblock ISSN 1365-2966.
\newblock \doi{10.1111/j.1365-2966.2003.07133.x}.
\newblock URL \url{http://dx.doi.org/10.1111/j.1365-2966.2003.07133.x}.

\bibitem[{Planck Collaboration} et~al.(2016){Planck Collaboration}, {Adam},
  {Ade}, {Aghanim}, {Akrami}, {Alves}, {Arg{\"u}eso}, {Arnaud}, {Arroja},
  {Ashdown}, {Aumont}, {Baccigalupi}, {Ballardini}, {Band ay}, {Barreiro},
  {Bartlett}, {Bartolo}, {Basak}, {Battaglia}, {Battaner}, {Battye}, {Benabed},
  {Beno{\^\i}t}, {Benoit-L{\'e}vy}, {Bernard}, {Bersanelli}, {Bertincourt},
  {Bielewicz}, {Bikmaev}, {Bock}, {B{\"o}hringer}, {Bonaldi}, {Bonavera},
  {Bond}, {Borrill}, {Bouchet}, {Boulanger}, {Bucher}, {Burenin}, {Burigana},
  {Butler}, {Calabrese}, {Cardoso}, {Carvalho}, {Casaponsa}, {Castex},
  {Catalano}, {Challinor}, {Chamballu}, {Chary}, {Chiang}, {Chluba}, {Chon},
  {Christensen}, {Church}, {Clemens}, {Clements}, {Colombi}, {Colombo},
  {Combet}, {Comis}, {Contreras}, {Couchot}, {Coulais}, {Crill}, {Cruz},
  {Curto}, {Cuttaia}, {Danese}, {Davies}, {Davis}, {de Bernardis}, {de Rosa},
  {de Zotti}, {Delabrouille}, {Delouis}, {D{\'e}sert}, {Di Valentino},
  {Dickinson}, {Diego}, {Dolag}, {Dole}, {Donzelli}, {Dor{\'e}}, {Douspis},
  {Ducout}, {Dunkley}, {Dupac}, {Efstathiou}, {Eisenhardt}, {Elsner},
  {En{\ss}lin}, {Eriksen}, {Falgarone}, {Fantaye}, {Farhang}, {Feeney},
  {Fergusson}, {Fernandez-Cobos}, {Feroz}, {Finelli}, {Florido}, {Forni},
  {Frailis}, {Fraisse}, {Franceschet}, {Franceschi}, {Frejsel}, {Frolov},
  {Galeotta}, {Galli}, {Ganga}, {Gauthier}, {G{\'e}nova-Santos}, {Gerbino},
  {Ghosh}, {Giard}, {Giraud-H{\'e}raud}, {Giusarma}, {Gjerl{\o}w},
  {Gonz{\'a}lez-Nuevo}, {G{\'o}rski}, {Grainge}, {Gratton}, {Gregorio},
  {Gruppuso}, {Gudmundsson}, {Hamann}, {Handley}, {Hansen}, {Hanson},
  {Harrison}, {Heavens}, {Helou}, {Henrot-Versill{\'e}},
  {Hern{\'a}ndez-Monteagudo}, {Herranz}, {Hildebrandt}, {Hivon}, {Hobson},
  {Holmes}, {Hornstrup}, {Hovest}, {Huang}, {Huffenberger}, {Hurier},
  {Ili{\'c}}, {Jaffe}, {Jaffe}, {Jin}, {Jones}, {Juvela}, {Karakci},
  {Keih{\"a}nen}, {Keskitalo}, {Khamitov}, {Kiiveri}, {Kim}, {Kisner},
  {Kneissl}, {Knoche}, {Knox}, {Krachmalnicoff}, {Kunz}, {Kurki-Suonio},
  {Lacasa}, {Lagache}, {L{\"a}hteenm{\"a}ki}, {Lamarre}, {Langer}, {Lasenby},
  {Lattanzi}, {Lawrence}, {Le Jeune}, {Leahy}, {Lellouch}, {Leonardi},
  {Le{\'o}n-Tavares}, {Lesgourgues}, {Levrier}, {Lewis}, {Liguori}, {Lilje},
  {Lilley}, {Linden-V{\o}rnle}, {Lindholm}, {Liu}, {L{\'o}pez-Caniego},
  {Lubin}, {Ma}, {Mac{\'\i}as-P{\'e}rez}, {Maggio}, {Maino}, {Mak},
  {Mandolesi}, {Mangilli}, {Marchini}, {Marcos-Caballero}, {Marinucci},
  {Maris}, {Marshall}, {Martin}, {Martinelli}, {Mart{\'\i}nez-Gonz{\'a}lez},
  {Masi}, {Matarrese}, {Mazzotta}, {McEwen}, {McGehee}, {Mei}, {Meinhold},
  {Melchiorri}, {Melin}, {Mendes}, {Mennella}, {Migliaccio}, {Mikkelsen},
  {Millea}, {Mitra}, {Miville-Desch{\^e}nes}, {Molinari}, {Moneti}, {Montier},
  {Moreno}, {Morgante}, {Mortlock}, {Moss}, {Mottet}, {M{\"u}nchmeyer},
  {Munshi}, {Murphy}, {Narimani}, {Naselsky}, {Nastasi}, {Nati}, {Natoli},
  {Negrello}, {Netterfield}, {N{\o}rgaard-Nielsen}, {Noviello}, {Novikov},
  {Novikov}, {Olamaie}, {Oppermann}, {Orlando}, {Oxborrow}, {Paci}, {Pagano},
  {Pajot}, {Paladini}, {Pandolfi}, {Paoletti}, {Partridge}, {Pasian},
  {Patanchon}, {Pearson}, {Peel}, {Peiris}, {Pelkonen}, {Perdereau}, {Perotto},
  {Perrott}, {Perrotta}, {Pettorino}, {Piacentini}, {Piat}, {Pierpaoli},
  {Pietrobon}, {Plaszczynski}, {Pogosyan}, {Pointecouteau}, {Polenta}, {Popa},
  {Pratt}, {Pr{\'e}zeau}, {Prunet}, {Puget}, {Rachen}, {Racine}, {Reach},
  {Rebolo}, {Reinecke}, {Remazeilles}, {Renault}, {Renzi}, {Ristorcelli},
  {Rocha}, {Roman}, {Romelli}, {Rosset}, {Rossetti}, {Rotti}, {Roudier},
  {Rouill{\'e} d'Orfeuil}, {Rowan-Robinson}, {Rubi{\~n}o-Mart{\'\i}n},
  {Ruiz-Granados}, {Rumsey}, {Rusholme}, {Said}, {Salvatelli}, {Salvati},
  {Sandri}, {Sanghera}, {Santos}, {Saunders}, {Sauv{\'e}}, {Savelainen},
  {Savini}, {Schaefer}, {Schammel}, {Scott}, {Seiffert}, {Serra}, {Shellard},
  {Shimwell}, {Shiraishi}, {Smith}, {Souradeep}, {Spencer}, {Spinelli},
  {Stanford}, {Stern}, {Stolyarov}, {Stompor}, {Strong}, {Sudiwala}, {Sunyaev},
  {Sutter}, {Sutton}, {Suur-Uski}, {Sygnet}, {Tauber}, {Tavagnacco}, {Terenzi},
  {Texier}, {Toffolatti}, {Tomasi}, {Tornikoski}, {Tramonte}, {Tristram},
  {Troja}, {Trombetti}, {Tucci}, {Tuovinen}, {T{\"u}rler}, {Umana},
  {Valenziano}, {Valiviita}, {Van Tent}, {Vassallo}, {Vibert}, {Vidal}, {Viel},
  {Vielva}, {Villa}, {Wade}, {Walter}, {Wand elt}, {Watson}, {Wehus},
  {Welikala}, {Weller}, {White}, {White}, {Wilkinson}, {Yvon}, {Zacchei},
  {Zibin}, and {Zonca}]{planck2016}
{Planck Collaboration}, R.~{Adam}, P.~A.~R. {Ade}, N.~{Aghanim}, Y.~{Akrami},
  M.~I.~R. {Alves}, F.~{Arg{\"u}eso}, M.~{Arnaud}, F.~{Arroja}, M.~{Ashdown},
  J.~{Aumont}, C.~{Baccigalupi}, M.~{Ballardini}, A.~J. {Band ay}, R.~B.
  {Barreiro}, J.~G. {Bartlett}, N.~{Bartolo}, S.~{Basak}, P.~{Battaglia},
  E.~{Battaner}, R.~{Battye}, K.~{Benabed}, A.~{Beno{\^\i}t},
  A.~{Benoit-L{\'e}vy}, J.~P. {Bernard}, M.~{Bersanelli}, B.~{Bertincourt},
  P.~{Bielewicz}, I.~{Bikmaev}, J.~J. {Bock}, H.~{B{\"o}hringer}, A.~{Bonaldi},
  L.~{Bonavera}, J.~R. {Bond}, J.~{Borrill}, F.~R. {Bouchet}, F.~{Boulanger},
  M.~{Bucher}, R.~{Burenin}, C.~{Burigana}, R.~C. {Butler}, E.~{Calabrese},
  J.~F. {Cardoso}, P.~{Carvalho}, B.~{Casaponsa}, G.~{Castex}, A.~{Catalano},
  A.~{Challinor}, A.~{Chamballu}, R.~R. {Chary}, H.~C. {Chiang}, J.~{Chluba},
  G.~{Chon}, P.~R. {Christensen}, S.~{Church}, M.~{Clemens}, D.~L. {Clements},
  S.~{Colombi}, L.~P.~L. {Colombo}, C.~{Combet}, B.~{Comis}, D.~{Contreras},
  F.~{Couchot}, A.~{Coulais}, B.~P. {Crill}, M.~{Cruz}, A.~{Curto},
  F.~{Cuttaia}, L.~{Danese}, R.~D. {Davies}, R.~J. {Davis}, P.~{de Bernardis},
  A.~{de Rosa}, G.~{de Zotti}, J.~{Delabrouille}, J.~M. {Delouis}, F.~X.
  {D{\'e}sert}, E.~{Di Valentino}, C.~{Dickinson}, J.~M. {Diego}, K.~{Dolag},
  H.~{Dole}, S.~{Donzelli}, O.~{Dor{\'e}}, M.~{Douspis}, A.~{Ducout},
  J.~{Dunkley}, X.~{Dupac}, G.~{Efstathiou}, P.~R.~M. {Eisenhardt},
  F.~{Elsner}, T.~A. {En{\ss}lin}, H.~K. {Eriksen}, E.~{Falgarone},
  Y.~{Fantaye}, M.~{Farhang}, S.~{Feeney}, J.~{Fergusson},
  R.~{Fernandez-Cobos}, F.~{Feroz}, F.~{Finelli}, E.~{Florido}, O.~{Forni},
  M.~{Frailis}, A.~A. {Fraisse}, C.~{Franceschet}, E.~{Franceschi},
  A.~{Frejsel}, A.~{Frolov}, S.~{Galeotta}, S.~{Galli}, K.~{Ganga},
  C.~{Gauthier}, R.~T. {G{\'e}nova-Santos}, M.~{Gerbino}, T.~{Ghosh},
  M.~{Giard}, Y.~{Giraud-H{\'e}raud}, E.~{Giusarma}, E.~{Gjerl{\o}w},
  J.~{Gonz{\'a}lez-Nuevo}, K.~M. {G{\'o}rski}, K.~J.~B. {Grainge},
  S.~{Gratton}, A.~{Gregorio}, A.~{Gruppuso}, J.~E. {Gudmundsson}, J.~{Hamann},
  W.~{Handley}, F.~K. {Hansen}, D.~{Hanson}, D.~L. {Harrison}, A.~{Heavens},
  G.~{Helou}, S.~{Henrot-Versill{\'e}}, C.~{Hern{\'a}ndez-Monteagudo},
  D.~{Herranz}, S.~R. {Hildebrandt}, E.~{Hivon}, M.~{Hobson}, W.~A. {Holmes},
  A.~{Hornstrup}, W.~{Hovest}, Z.~{Huang}, K.~M. {Huffenberger}, G.~{Hurier},
  S.~{Ili{\'c}}, A.~H. {Jaffe}, T.~R. {Jaffe}, T.~{Jin}, W.~C. {Jones},
  M.~{Juvela}, A.~{Karakci}, E.~{Keih{\"a}nen}, R.~{Keskitalo}, I.~{Khamitov},
  K.~{Kiiveri}, J.~{Kim}, T.~S. {Kisner}, R.~{Kneissl}, J.~{Knoche}, L.~{Knox},
  N.~{Krachmalnicoff}, M.~{Kunz}, H.~{Kurki-Suonio}, F.~{Lacasa}, G.~{Lagache},
  A.~{L{\"a}hteenm{\"a}ki}, J.~M. {Lamarre}, M.~{Langer}, A.~{Lasenby},
  M.~{Lattanzi}, C.~R. {Lawrence}, M.~{Le Jeune}, J.~P. {Leahy}, E.~{Lellouch},
  R.~{Leonardi}, J.~{Le{\'o}n-Tavares}, J.~{Lesgourgues}, F.~{Levrier},
  A.~{Lewis}, M.~{Liguori}, P.~B. {Lilje}, M.~{Lilley}, M.~{Linden-V{\o}rnle},
  V.~{Lindholm}, H.~{Liu}, M.~{L{\'o}pez-Caniego}, P.~M. {Lubin}, Y.~Z. {Ma},
  J.~F. {Mac{\'\i}as-P{\'e}rez}, G.~{Maggio}, D.~{Maino}, D.~S.~Y. {Mak},
  N.~{Mandolesi}, A.~{Mangilli}, A.~{Marchini}, A.~{Marcos-Caballero},
  D.~{Marinucci}, M.~{Maris}, D.~J. {Marshall}, P.~G. {Martin},
  M.~{Martinelli}, E.~{Mart{\'\i}nez-Gonz{\'a}lez}, S.~{Masi}, S.~{Matarrese},
  P.~{Mazzotta}, J.~D. {McEwen}, P.~{McGehee}, S.~{Mei}, P.~R. {Meinhold},
  A.~{Melchiorri}, J.~B. {Melin}, L.~{Mendes}, A.~{Mennella}, M.~{Migliaccio},
  K.~{Mikkelsen}, M.~{Millea}, S.~{Mitra}, M.~A. {Miville-Desch{\^e}nes},
  D.~{Molinari}, A.~{Moneti}, L.~{Montier}, R.~{Moreno}, G.~{Morgante},
  D.~{Mortlock}, A.~{Moss}, S.~{Mottet}, M.~{M{\"u}nchmeyer}, D.~{Munshi},
  J.~A. {Murphy}, A.~{Narimani}, P.~{Naselsky}, A.~{Nastasi}, F.~{Nati},
  P.~{Natoli}, M.~{Negrello}, C.~B. {Netterfield}, H.~U. {N{\o}rgaard-Nielsen},
  F.~{Noviello}, D.~{Novikov}, I.~{Novikov}, M.~{Olamaie}, N.~{Oppermann},
  E.~{Orlando}, C.~A. {Oxborrow}, F.~{Paci}, L.~{Pagano}, F.~{Pajot},
  R.~{Paladini}, S.~{Pandolfi}, D.~{Paoletti}, B.~{Partridge}, F.~{Pasian},
  G.~{Patanchon}, T.~J. {Pearson}, M.~{Peel}, H.~V. {Peiris}, V.~M. {Pelkonen},
  O.~{Perdereau}, L.~{Perotto}, Y.~C. {Perrott}, F.~{Perrotta}, V.~{Pettorino},
  F.~{Piacentini}, M.~{Piat}, E.~{Pierpaoli}, D.~{Pietrobon},
  S.~{Plaszczynski}, D.~{Pogosyan}, E.~{Pointecouteau}, G.~{Polenta},
  L.~{Popa}, G.~W. {Pratt}, G.~{Pr{\'e}zeau}, S.~{Prunet}, J.~L. {Puget}, J.~P.
  {Rachen}, B.~{Racine}, W.~T. {Reach}, R.~{Rebolo}, M.~{Reinecke},
  M.~{Remazeilles}, C.~{Renault}, A.~{Renzi}, I.~{Ristorcelli}, G.~{Rocha},
  M.~{Roman}, E.~{Romelli}, C.~{Rosset}, M.~{Rossetti}, A.~{Rotti},
  G.~{Roudier}, B.~{Rouill{\'e} d'Orfeuil}, M.~{Rowan-Robinson}, J.~A.
  {Rubi{\~n}o-Mart{\'\i}n}, B.~{Ruiz-Granados}, C.~{Rumsey}, B.~{Rusholme},
  N.~{Said}, V.~{Salvatelli}, L.~{Salvati}, M.~{Sandri}, H.~S. {Sanghera},
  D.~{Santos}, R.~D.~E. {Saunders}, A.~{Sauv{\'e}}, M.~{Savelainen},
  G.~{Savini}, B.~M. {Schaefer}, M.~P. {Schammel}, D.~{Scott}, M.~D.
  {Seiffert}, P.~{Serra}, E.~P.~S. {Shellard}, T.~W. {Shimwell},
  M.~{Shiraishi}, K.~{Smith}, T.~{Souradeep}, L.~D. {Spencer}, M.~{Spinelli},
  S.~A. {Stanford}, D.~{Stern}, V.~{Stolyarov}, R.~{Stompor}, A.~W. {Strong},
  R.~{Sudiwala}, R.~{Sunyaev}, P.~{Sutter}, D.~{Sutton}, A.~S. {Suur-Uski},
  J.~F. {Sygnet}, J.~A. {Tauber}, D.~{Tavagnacco}, L.~{Terenzi}, D.~{Texier},
  L.~{Toffolatti}, M.~{Tomasi}, M.~{Tornikoski}, D.~{Tramonte}, M.~{Tristram},
  A.~{Troja}, T.~{Trombetti}, M.~{Tucci}, J.~{Tuovinen}, M.~{T{\"u}rler},
  G.~{Umana}, L.~{Valenziano}, J.~{Valiviita}, F.~{Van Tent}, T.~{Vassallo},
  L.~{Vibert}, M.~{Vidal}, M.~{Viel}, P.~{Vielva}, F.~{Villa}, L.~A. {Wade},
  B.~{Walter}, B.~D. {Wand elt}, R.~{Watson}, I.~K. {Wehus}, N.~{Welikala},
  J.~{Weller}, M.~{White}, S.~D.~M. {White}, A.~{Wilkinson}, D.~{Yvon},
  A.~{Zacchei}, J.~P. {Zibin}, and A.~{Zonca}.
\newblock {Planck 2015 results. I. Overview of products and scientific
  results}.
\newblock \emph{\aap}, 594:\penalty0 A1, September 2016.
\newblock \doi{10.1051/0004-6361/201527101}.

\bibitem[Pritchard and Loeb(2012)]{PritchardLoeb12}
Jonathan~R Pritchard and Abraham Loeb.
\newblock 21 cm cosmology in the 21st century.
\newblock \emph{Reports on Progress in Physics}, 75\penalty0 (8):\penalty0
  086901, Jul 2012.
\newblock ISSN 1361-6633.
\newblock \doi{10.1088/0034-4885/75/8/086901}.
\newblock URL \url{http://dx.doi.org/10.1088/0034-4885/75/8/086901}.

\bibitem[Ronneberger et~al.(2015)Ronneberger, Fischer, and Brox]{unet_review}
Olaf Ronneberger, Philipp Fischer, and Thomas Brox.
\newblock U-net: Convolutional networks for biomedical image segmentation.
\newblock \emph{CoRR}, abs/1505.04597, 2015.
\newblock URL \url{http://arxiv.org/abs/1505.04597}.

\bibitem[{Rybicki} and {Lightman}(1986)]{RybickLightman86}
George~B. {Rybicki} and Alan~P. {Lightman}.
\newblock \emph{{Radiative Processes in Astrophysics}}.
\newblock 1986.

\bibitem[{Santos} et~al.(2005){Santos}, {Cooray}, and {Knox}]{santos-fg}
M{\'a}rio~G. {Santos}, Asantha {Cooray}, and Lloyd {Knox}.
\newblock {Multifrequency Analysis of 21 Centimeter Fluctuations from the Era
  of Reionization}.
\newblock \emph{\apj}, 625\penalty0 (2):\penalty0 575--587, June 2005.
\newblock \doi{10.1086/429857}.

\bibitem[{Schnitzeler} et~al.(2009){Schnitzeler}, {Katgert}, and {de
  Bruyn}]{schnitzeler2009}
D.~H.~F.~M. {Schnitzeler}, P.~{Katgert}, and A.~G. {de Bruyn}.
\newblock {WSRT Faraday tomography of the Galactic ISM at
  {\ensuremath{\lambda}} \raisebox{-0.5ex}\textasciitilde 0.86 m. I. The GEMINI
  data set at (l, b) = (181{\textdegree}, 20{\textdegree})}.
\newblock \emph{\aap}, 494\penalty0 (2):\penalty0 611--622, February 2009.
\newblock \doi{10.1051/0004-6361:20078912}.

\bibitem[{Scott} and {Rees}(1990)]{scott-dark-ages}
D.~{Scott} and M.~J. {Rees}.
\newblock {The 21-cm line at high redshift: a diagnostic for the origin of
  large scale structure}.
\newblock \emph{\mnras}, 247:\penalty0 510, December 1990.

\bibitem[Sekiguchi et~al.(2019)Sekiguchi, Takahashi, Tashiro, and
  Yokoyama]{gaussSekiguchi_2019}
Toyokazu Sekiguchi, Tomo Takahashi, Hiroyuki Tashiro, and Shuichiro Yokoyama.
\newblock Probing primordial non-gaussianity with 21 cm fluctuations from
  minihalos.
\newblock \emph{Journal of Cosmology and Astroparticle Physics}, 2019\penalty0
  (02):\penalty0 033–033, Feb 2019.
\newblock ISSN 1475-7516.
\newblock \doi{10.1088/1475-7516/2019/02/033}.
\newblock URL \url{http://dx.doi.org/10.1088/1475-7516/2019/02/033}.

\bibitem[Shaw et~al.(2014)Shaw, Sigurdson, Pen, Stebbins, and
  Sitwell]{Shaw_2014}
J.~Richard Shaw, Kris Sigurdson, Ue-Li Pen, Albert Stebbins, and Michael
  Sitwell.
\newblock All-sky interferometry with spherical harmonic transit telescopes.
\newblock \emph{The Astrophysical Journal}, 781\penalty0 (2):\penalty0 57, Jan
  2014.
\newblock ISSN 1538-4357.
\newblock \doi{10.1088/0004-637x/781/2/57}.
\newblock URL \url{http://dx.doi.org/10.1088/0004-637X/781/2/57}.

\bibitem[Shaw et~al.(2015{\natexlab{a}})Shaw, Sigurdson, Sitwell, Stebbins, and
  Pen]{Shaw_2015}
J.~Richard Shaw, Kris Sigurdson, Michael Sitwell, Albert Stebbins, and Ue-Li
  Pen.
\newblock Coaxing cosmic 21 cm fluctuations from the polarized sky usingm-mode
  analysis.
\newblock \emph{Physical Review D}, 91\penalty0 (8), Apr 2015{\natexlab{a}}.
\newblock ISSN 1550-2368.
\newblock \doi{10.1103/physrevd.91.083514}.
\newblock URL \url{http://dx.doi.org/10.1103/PhysRevD.91.083514}.

\bibitem[Shaw et~al.(2015{\natexlab{b}})Shaw, Sigurdson, Sitwell, Stebbins, and
  Pen]{polar-Shaw_2015}
J.~Richard Shaw, Kris Sigurdson, Michael Sitwell, Albert Stebbins, and Ue-Li
  Pen.
\newblock Coaxing cosmic 21 cm fluctuations from the polarized sky usingm-mode
  analysis.
\newblock \emph{Physical Review D}, 91\penalty0 (8), Apr 2015{\natexlab{b}}.
\newblock ISSN 1550-2368.
\newblock \doi{10.1103/physrevd.91.083514}.
\newblock URL \url{http://dx.doi.org/10.1103/PhysRevD.91.083514}.

\bibitem[{Slosar} et~al.(2019){Slosar}, {Ahmed}, {Alonso}, {Amin}, {Arena},
  {Bandura}, {Battaglia}, {Blazek}, {Bull}, {Castorina}, {Chang}, {Connor},
  {Dav{\'e}}, {Dvorkin}, {van Engelen}, {Ferraro}, {Flauger}, {Foreman},
  {Frisch}, {Green}, {Holder}, {Jacobs}, {Johnson}, {Dillon}, {Karagiannis},
  {Kaurov}, {Knox}, {Liu}, {Loverde}, {Ma}, {Masui}, {McClintock}, {Moodley},
  {Munchmeyer}, {Newburgh}, {Ng}, {Nomerotski}, {O'Connor}, {Obuljen},
  {Padmanabhan}, {Parkinson}, {Prochaska}, {Rajendran}, {Rapetti},
  {Saliwanchik}, {Schaan}, {Sehgal}, {Shaw}, {Sheehy}, {Sheldon}, {Shirley},
  {Silverstein}, {Slatyer}, {Slosar}, {Stankus}, {Stebbins}, {Timbie},
  {Tucker}, {Tyndall}, {Villaescusa Navarro}, {Wallisch}, and {White}]{PUMA}
Anze {Slosar}, Zeeshan {Ahmed}, David {Alonso}, Mustafa~A. {Amin}, Evan~J.
  {Arena}, Kevin {Bandura}, Nicholas {Battaglia}, Jonathan {Blazek}, Philip
  {Bull}, Emanuele {Castorina}, Tzu-Ching {Chang}, Liam {Connor}, Romeel
  {Dav{\'e}}, Cora {Dvorkin}, Alexander {van Engelen}, Simone {Ferraro},
  Raphael {Flauger}, Simon {Foreman}, Josef {Frisch}, Daniel {Green}, Gilbert
  {Holder}, Daniel {Jacobs}, Matthew~C. {Johnson}, Joshua~S. {Dillon},
  Dionysios {Karagiannis}, Alexander~A. {Kaurov}, Lloyd {Knox}, Adrian {Liu},
  Marilena {Loverde}, Yin-Zhe {Ma}, Kiyoshi~W. {Masui}, Thomas {McClintock},
  Kavilan {Moodley}, Moritz {Munchmeyer}, Laura~B. {Newburgh}, Cherry {Ng},
  Andrei {Nomerotski}, Paul {O'Connor}, Andrej {Obuljen}, Hamsa {Padmanabhan},
  David {Parkinson}, J.~Xavier {Prochaska}, Surjeet {Rajendran}, David
  {Rapetti}, Benjamin {Saliwanchik}, Emmanuel {Schaan}, Neelima {Sehgal},
  J.~Richard {Shaw}, Chris {Sheehy}, Erin {Sheldon}, Raphael {Shirley}, Eva
  {Silverstein}, Tracy {Slatyer}, Anze {Slosar}, Paul {Stankus}, Albert
  {Stebbins}, Peter~T. {Timbie}, Gregory~S. {Tucker}, William {Tyndall},
  Francisco {Villaescusa Navarro}, Benjamin {Wallisch}, and Martin {White}.
\newblock {Packed Ultra-wideband Mapping Array (PUMA): A Radio Telescope for
  Cosmology and Transients}.
\newblock In \emph{Bulletin of the American Astronomical Society}, volume~51,
  page~53, September 2019.

\bibitem[Spinelli et~al.(2019)Spinelli, Bernardi, and Santos]{Spinelli_2019}
Marta Spinelli, Gianni Bernardi, and Mario~G Santos.
\newblock On the contamination of the global 21 cm signal from polarized
  foregrounds.
\newblock \emph{Monthly Notices of the Royal Astronomical Society}, Sep 2019.
\newblock ISSN 1365-2966.
\newblock \doi{10.1093/mnras/stz2425}.
\newblock URL \url{http://dx.doi.org/10.1093/mnras/stz2425}.

\bibitem[{Switzer} et~al.(2015){Switzer}, {Chang}, {Masui}, {Pen}, and
  {Voytek}]{2015ApJ_signal_loss_singledish}
E.~R. {Switzer}, T.~C. {Chang}, K.~W. {Masui}, U.~L. {Pen}, and T.~C. {Voytek}.
\newblock {Interpreting the Unresolved Intensity of Cosmologically Redshifted
  Line Radiation}.
\newblock \emph{\apj}, 815\penalty0 (1):\penalty0 51, December 2015.
\newblock \doi{10.1088/0004-637X/815/1/51}.

\bibitem[Switzer et~al.(2015)Switzer, Chang, Masui, Pen, and
  Voytek]{Switzer_2015_pca_real}
E.~R. Switzer, T.-C. Chang, K.~W. Masui, U.-L. Pen, and T.~C. Voytek.
\newblock {INTERPRETING} {THE} {UNRESOLVED} {INTENSITY} {OF} {COSMOLOGICALLY}
  {REDSHIFTED} {LINE} {RADIATION}.
\newblock \emph{The Astrophysical Journal}, 815\penalty0 (1):\penalty0 51, dec
  2015.
\newblock \doi{10.1088/0004-637x/815/1/51}.
\newblock URL \url{https://doi.org/10.1088%2F0004-637x%2F815%2F1%2F51}.

\bibitem[Tegmark et~al.(2000)Tegmark, Eisenstein, Hu, and
  de~Oliveira-Costa]{Tegmark_2000}
Max Tegmark, Daniel~J. Eisenstein, Wayne Hu, and Angelica de~Oliveira-Costa.
\newblock Foregrounds and forecasts for the cosmic microwave background.
\newblock \emph{The Astrophysical Journal}, 530\penalty0 (1):\penalty0
  133--165, feb 2000.
\newblock \doi{10.1086/308348}.
\newblock URL \url{https://doi.org/10.1086%2F308348}.

\bibitem[Tipping and Bishop(1999)]{bishop_prob_pca}
Michael~E. Tipping and Christopher~M. Bishop.
\newblock Probabilistic principal component analysis.
\newblock \emph{Journal of the Royal Statistical Society. Series B (Statistical
  Methodology)}, 61\penalty0 (3):\penalty0 611--622, 1999.
\newblock ISSN 13697412, 14679868.
\newblock URL \url{http://www.jstor.org/stable/2680726}.

\bibitem[{Tozzi} et~al.(2000){Tozzi}, {Madau}, {Meiksin}, and {Rees}]{Tozzi00}
Paolo {Tozzi}, Piero {Madau}, Avery {Meiksin}, and Martin~J. {Rees}.
\newblock {Radio Signatures of H I at High Redshift: Mapping the End of the
  ``Dark Ages''}.
\newblock \emph{\apj}, 528\penalty0 (2):\penalty0 597--606, January 2000.
\newblock \doi{10.1086/308196}.

\bibitem[Villaescusa-Navarro et~al.(2014)Villaescusa-Navarro, Viel, Alonso,
  Datta, Bull, and Santos]{villaescusanavarro2014crosscorrelating}
Francisco Villaescusa-Navarro, Matteo Viel, David Alonso, Kanan~K. Datta,
  Philip Bull, and Mario~G. Santos.
\newblock Cross-correlating 21cm intensity maps with lyman break galaxies in
  the post-reionization era, 2014.

\bibitem[{Villaescusa-Navarro} et~al.(2017){Villaescusa-Navarro}, {Alonso}, and
  {Viel}]{paco-2016}
Francisco {Villaescusa-Navarro}, David {Alonso}, and Matteo {Viel}.
\newblock {Baryonic acoustic oscillations from 21 cm intensity mapping: the
  Square Kilometre Array case}.
\newblock \emph{\mnras}, 466\penalty0 (3):\penalty0 2736--2751, April 2017.
\newblock \doi{10.1093/mnras/stw3224}.

\bibitem[{Villaescusa-Navarro} et~al.(2018){Villaescusa-Navarro}, {Genel},
  {Castorina}, {Obuljen}, {Spergel}, {Hernquist}, {Nelson}, {Carucci},
  {Pillepich}, {Marinacci}, {Diemer}, {Vogelsberger}, {Weinberger}, and
  {Pakmor}]{Paco_2018}
Francisco {Villaescusa-Navarro}, Shy {Genel}, Emanuele {Castorina}, Andrej
  {Obuljen}, David~N. {Spergel}, Lars {Hernquist}, Dylan {Nelson}, Isabella~P.
  {Carucci}, Annalisa {Pillepich}, Federico {Marinacci}, Benedikt {Diemer},
  Mark {Vogelsberger}, Rainer {Weinberger}, and R{\"u}diger {Pakmor}.
\newblock {Ingredients for 21 cm Intensity Mapping}.
\newblock \emph{\apj}, 866\penalty0 (2):\penalty0 135, October 2018.
\newblock \doi{10.3847/1538-4357/aadba0}.

\bibitem[{Villanueva-Domingo} and {Villaescusa-Navarro}(2020)]{Pablo_2020}
Pablo {Villanueva-Domingo} and Francisco {Villaescusa-Navarro}.
\newblock {Removing Astrophysics in 21 cm maps with Neural Networks}.
\newblock \emph{arXiv e-prints}, art. arXiv:2006.14305, June 2020.

\bibitem[{Wadekar} et~al.(2020){Wadekar}, {Villaescusa-Navarro}, {Ho}, and
  {Perreault-Levasseur}]{Jay_2020}
Digvijay {Wadekar}, Francisco {Villaescusa-Navarro}, Shirley {Ho}, and Laurence
  {Perreault-Levasseur}.
\newblock {HInet: Generating neutral hydrogen from dark matter with neural
  networks}.
\newblock \emph{arXiv e-prints}, art. arXiv:2007.10340, July 2020.

\bibitem[{Waelkens} et~al.(2009){Waelkens}, {Jaffe}, {Reinecke}, {Kitaura}, and
  {En{\ss}lin}]{hammurabi-gal-fg}
A.~{Waelkens}, T.~{Jaffe}, M.~{Reinecke}, F.~S. {Kitaura}, and T.~A.
  {En{\ss}lin}.
\newblock {Simulating polarized Galactic synchrotron emission at all
  frequencies. The Hammurabi code}.
\newblock \emph{\aap}, 495\penalty0 (2):\penalty0 697--706, February 2009.
\newblock \doi{10.1051/0004-6361:200810564}.

\bibitem[{Wang} et~al.(2006){Wang}, {Tegmark}, {Santos}, and
  {Knox}]{wang_poly_2006}
Xiaomin {Wang}, Max {Tegmark}, M{\'a}rio~G. {Santos}, and Lloyd {Knox}.
\newblock {21 cm Tomography with Foregrounds}.
\newblock \emph{\apj}, 650\penalty0 (2):\penalty0 529--537, October 2006.
\newblock \doi{10.1086/506597}.

\bibitem[Weltman et~al.(2020)Weltman, Bull, Camera, Kelley, Padmanabhan,
  Pritchard, Raccanelli, Riemer-Sørensen, Shao, Andrianomena, and
  et~al.]{Weltman_2020}
A.~Weltman, P.~Bull, S.~Camera, K.~Kelley, H.~Padmanabhan, J.~Pritchard,
  A.~Raccanelli, S.~Riemer-Sørensen, L.~Shao, S.~Andrianomena, and et~al.
\newblock Fundamental physics with the square kilometre array.
\newblock \emph{Publications of the Astronomical Society of Australia}, 37,
  2020.
\newblock ISSN 1448-6083.
\newblock \doi{10.1017/pasa.2019.42}.
\newblock URL \url{http://dx.doi.org/10.1017/pasa.2019.42}.

\bibitem[Wilson(2011)]{wilson2011techniques}
T.~L. Wilson.
\newblock Techniques of radio astronomy, 2011.

\bibitem[{Wolleben} et~al.(2006){Wolleben}, {Landecker}, {Reich}, and
  {Wielebinski}]{wolleben-polar2006A&A...448..411W}
M.~{Wolleben}, T.~L. {Landecker}, W.~{Reich}, and R.~{Wielebinski}.
\newblock {An absolutely calibrated survey of polarized emission from the
  northern sky at 1.4 GHz. Observations and data reduction}.
\newblock \emph{\aap}, 448\penalty0 (1):\penalty0 411--424, March 2006.
\newblock \doi{10.1051/0004-6361:20053851}.

\bibitem[Wolz et~al.(2014)Wolz, Abdalla, Blake, Shaw, Chapman, and
  Rawlings]{Wolz_2014}
L.~Wolz, F.~B. Abdalla, C.~Blake, J.~R. Shaw, E.~Chapman, and S.~Rawlings.
\newblock The effect of foreground subtraction on cosmological measurements
  from intensity mapping.
\newblock \emph{Monthly Notices of the Royal Astronomical Society},
  441\penalty0 (4):\penalty0 3271–3283, May 2014.
\newblock ISSN 0035-8711.
\newblock \doi{10.1093/mnras/stu792}.
\newblock URL \url{http://dx.doi.org/10.1093/mnras/stu792}.

\bibitem[Wyithe et~al.(2007)Wyithe, Loeb, and Geil]{Wyithe_2007}
J.~Stuart~B. Wyithe, Abraham Loeb, and Paul~M. Geil.
\newblock Baryonic acoustic oscillations in 21-cm emission: a probe of dark
  energy out to high redshifts.
\newblock \emph{Monthly Notices of the Royal Astronomical Society},
  383\penalty0 (3):\penalty0 1195–1209, Dec 2007.
\newblock ISSN 1365-2966.
\newblock \doi{10.1111/j.1365-2966.2007.12631.x}.
\newblock URL \url{http://dx.doi.org/10.1111/j.1365-2966.2007.12631.x}.

\bibitem[Yao(2018)]{yao_2018}
Jian Yao.
\newblock \emph{International Symposium on Cosmology and Ali CMB Polarization
  Telescope}.
\newblock Sep 2018.
\newblock URL
  \url{https://indico.leeinst.sjtu.edu.cn/event/44/attachments/128/345/Foreground_removal_GAN.pdf}.

\bibitem[{Zhang} et~al.(2016){Zhang}, {Bunn}, {Karakci}, {Korotkov}, {Sutter},
  {Timbie}, {Tucker}, and {Wandelt}]{bayesian-semi-blind}
Le~{Zhang}, Emory~F. {Bunn}, Ata {Karakci}, Andrei {Korotkov}, P.~M. {Sutter},
  Peter~T. {Timbie}, Gregory~S. {Tucker}, and Benjamin~D. {Wandelt}.
\newblock {Bayesian Semi-blind Component Separation for Foreground Removal in
  Interferometric 21 cm Observations}.
\newblock \emph{\apjs}, 222\penalty0 (1):\penalty0 3, January 2016.
\newblock \doi{10.3847/0067-0049/222/1/3}.

\bibitem[Zonca et~al.(2019)Zonca, Singer, Lenz, Reinecke, Rosset, Hivon, and
  Gorski]{healpy}
Andrea Zonca, Leo Singer, Daniel Lenz, Martin Reinecke, Cyrille Rosset, Eric
  Hivon, and Krzysztof Gorski.
\newblock healpy: equal area pixelization and spherical harmonics transforms
  for data on the sphere in python.
\newblock \emph{Journal of Open Source Software}, 4\penalty0 (35):\penalty0
  1298, March 2019.
\newblock \doi{10.21105/joss.01298}.
\newblock URL \url{https://doi.org/10.21105/joss.01298}.

\end{thebibliography}



\appendix
\section{Hyperparameter choices}
\label{sec:appendix}

\begin{table}[h!]
\adjustbox{width=\textwidth, center}{
 \begin{tabular}{l l l p{6.75cm}}
 \toprule
 \textbf{Symbol} & \textbf{Prior Distribution} & \textbf{Optimum} & \textbf{Description} \\
 \midrule
\texttt{conv ND} & [2,3] & 3 & convolution filter dimensions \\
 \texttt{h} & disc $\mathcal{U}(2, 6)$& 6 & no. of down-convolutions  \\
 \texttt{w} & disc $\mathcal{U}(1, 6)$ & 3 & no. of convolutions for each conv. block \\
 \texttt{batchnorm} & [0,1] & \texttt{True}$^*$ & batch normalization for given layer  \\
 \texttt{batch size} & $\log \mathcal{U}(4, 24)$ & 48 & no. of samples per gradient descent step\\
 \texttt{n}$_{\rm filters}$ & [8,16,32] & 32 & initial number of conv. filters  \\
 $\beta_{\rm mom}$ & $\log\mathcal{U}(0.001, 0.75)$ & 0.05 & batch normalization momentum  \\
 $\lambda$ & $\log \mathcal{N}(-8.5, 2.5)$ & 0.0002 & learning rate for \texttt{Adam} optimizer  \\
 \toprule
\end{tabular}
}
\caption{Table of UNet parameters varied in architecture design, with optimal values (per GPU processor) shown. The algorithm \texttt{HyperOpt} was used in conjunction with the \texttt{Adam} optimizer tuning according to \cite{hyperopt}. The discrete uniform distribution is denoted by disc $\mathcal{U}(\cdot)$.  $^*${Batch normalization adopted for encoder layers}.}
\label{tab:hypers}
\end{table}

\end{document}